\documentstyle[amssymb,multicol,epsfig,prb,aps]{revtex}

\begin{document}
\draft

\title{Nucleation on top of islands in epitaxial growth}

\author{Stefan Heinrichs, J\"org Rottler{\footnote{Current address: 
        Department of Physics and Astronomy, 
        The Johns Hopkins University,\\\hspace*{0.3cm} 3400 N.~Charles Street,
        Baltimore, MD 21218, U.S.A.}}, and Philipp Maass}

\address{Fachbereich Physik, Universit\"at Konstanz, 
         D-78457 Konstanz,Germany}

\date{March 31, 2000}

\maketitle

\begin{abstract}
  We develop a theory for nucleation on top of islands in epitaxial
  growth based on the derivation of lifetimes and rates governing
  individual microscopic processes. These in particular include the
  encounter rate of $j$ atoms in a state, where in total $n\!\ge\!j$
  atoms are present on top of the island, and the lifetime of this
  state. The latter depends strongly on the additional step edge
  barrier $\Delta E_{\rm S}$ for descending atoms. We present two
  analytical approaches complemented by kinetic Monte Carlo
  simulations. In the first approach we employ a simplified stochastic
  description, which allows us to derive the nucleation rate on top of
  islands explicitly, if the dissociation times of unstable clusters
  can be neglected. We find that for small critical nuclei of size
  $i\!\le\!2$ the nucleation is governed by fluctuations, during which
  by chance $i\!+\!1$ atoms are present on the island. For large
  critical nuclei $i\!\ge\!3$ by contrast, the nucleation process can
  be described in a mean-field type manner, which for large $\Delta
  E_{\rm S}$ corresponds to the approach developed by Tersoff {\it et
    al.}  [Phys.~Rev.~Lett.\ {\bf 72}, 266 (1994)]. In both the
  fluctuation-dominated and the mean-field case, various scaling
  regimes are identified, where the typical island size at the onset
  of nucleation shows a power law in dependence on the adatom
  diffusion rates, the incoming atom flux, and the step edge crossing
  probability $\exp(-\Delta E_{\rm S}/k_{\rm B}T)$. Although it is
  possible to extend the simplified approach to more general
  situations, its applicability is limited, if dissociation rates of
  metastable clusters enter the problem as additional parameters. For
  such situations the second semi-analytical approach becomes
  superior.  This approach is based on novel rate equations , which
  can easily be solved numerically. Both theoretical approaches yield
  good agreement with Monte Carlo data. Implications for various
  applications are pointed out.
\end{abstract}

\begin{multicols}{2}\narrowtext

\section{Introduction}\label{intro-sec}

A fundamental problem in the theory of thin film growth
\cite{Venables/etal:1984} is the question, under which condition flat,
two-dimensional films form on the substrate surface in contrast to
mutually-separated, three-dimensional clusters. For films growing
under equilibrium conditions, this question has been answered many
years ago: If the interfacial tension between the substrate and
adsorbate is larger than the difference of the respective surface free
energies, then cluster formation is preferred (``Volmer-Weber growth''
\cite{Volmer/Weber:1926}), while a smaller (or equal) interfacial
tension leads to the formation of flat films (``Van-der Merwe
growth'').  An intermediate case is the ``Stranski-Krastanov growth''
mode,\cite{Stranski/Krastanov:1938} where cluster formation sets in
after the thickness of an initially smooth film exceeds a critical
height. This case may be understood from an interfacial tension that
varies with the film thickness. More recently, the influence of strain
effects on equilibrium film morphologies has been investigated by
various authors.\cite{Bauer/Merve:1986}

Films developing in the process of molecular beam epitaxy (MBE) are
usually not in thermodynamic equilibrium.\cite{Brune:1998}. Rapid
growth of films is achieved by a high supersaturation of the vapor,
and growth kinetics is governed by evaporation, diffusion and
aggregation processes far from equilibrium. Determining the film
morphology in this situation is a problem of stochastic dynamics. (For
a recent review on both kinetically and thermodynamically induced
instabilities in MBE see ref.~\onlinecite{Politi/etal:2000}.) During
the MBE experiment, two-dimensional islands composed of adsorbate
atoms form on the substrate. If these islands coalesce before stable
clusters nucleate on top of the islands in the second layer, a flat
two-dimensional film results. By contrast, if the onset of second
layer nucleation precedes island coalescence, then three-dimensional
cluster formation is obtained. The term ``second layer nucleation''
should not be taken literally here but rather should apply to the
formation of stable nuclei composed of $(i\!+\!1)$ atoms on top of
islands in general. As for the equilibrium structures, it might be
possible that cluster formation sets in above a certain film
thickness, when the relevant parameters governing the nucleation of
stable clusters on top of islands (see below) depend sensitively on
the film thickness. However, despite this similarity of the possible
growth processes with the equilibrium growth modes, it should be noted
that the dynamic problem is very different. In MBE flat films can be
produced even if the adsorbate does not wet the
substrate.\cite{Brune:1998}

The first theory for second layer nucleation in MBE was set up by
J.~Tersoff, A.~W.~Denier van der Gon, and
R.~M.~Tromp,\cite{Tersoff/etal:1994} which will be referred to as
``TDT approach'' in the following. Solving the stationary diffusion
equation in the presence of an incoming flux and employing classical
nucleation theory,\cite{Venables/etal:1984} these authors succeeded in
deriving an explicit expression for the rate of nucleation $\Omega(R)$
on top of circular shaped islands of radius $R$.  Assuming all island
radii to evolve approximately as the mean island radius $R(t)\sim
t^{1/2}$ at time $t$ (this situation will be referred to as the
``single-island model'' in the following), they calculated the
fraction $f_0(t)$ of ``covered islands'' (i.e.\ on top of which a
stable cluster has nucleated) from $\Omega(R)$. It turned out that
$f_0(t)$ rises from zero to one in the vicinity of a ``critical time''
$t_c$, which allows one to define a critical island radius
$R_c\!\equiv\!R(t_c)$ for second layer nucleation.  A simple criterion
for the occurrence of ``rough multilayer'' as opposed to smooth
``layer-by-layer growth'' is that $R_c$ is smaller than the mean
distance $l$ between islands in the first layer.

An important factor controlling the film morphology is the additional
step edge barrier $\Delta E_{\rm S}\!=\!E_{\rm S}\!-\!E_0$
(Ehrlich-Schwoebel barrier\cite{Ehrlich/Hudda:1966+Schwoebel:1969})
that has to be surmounted by an adatom in addition to the bare surface
diffusion barrier $E_0$, when an adatom crosses an island
edge.\cite{Zhang+Lagally:1994} For larger $\Delta E_{\rm S}$ one
expects adatoms to remain longer on islands and therefore to
accumulate more easily, which would lead to an increased second layer
nucleation rate $\Omega(R)$ and a smaller $R_c$. In fact, the theory
predicts that only for sufficiently large $\Delta E_{\rm S}$
three-dimensional clusters can occur on the substrate (for an
alternative possibility see however
ref.~\onlinecite{DasSarma/etal:1999}). Using the TDT approach,
$\Delta E_{\rm S}$ was estimated for a variety of different systems.
\cite{Bromann/etal:1995,Meyer/etal:1995,Smilauer/Harris:1995,Markov:1996,Roos+Tringides:1998}

An alternative approach for treating the problem of second layer
nucleation within a stochastic description based on scaling arguments
was developed recently by us.\cite{Rottler/Maass:1999} It was shown
that for $i\!=\!1$ the TDT approach is not applicable, but the
detailed treatment of fluctuations with only two atoms on top of the
island yields a correct description of the process (see also
ref.\onlinecite{Krug/etal:2000}; for an earlier approach focusing on
one dimension see ref.~\onlinecite{Elkinani/Villain:1994}). In this
work we will extend our former study of second layer nucleation by
means of both kinetic Monte Carlo simulations and scaling analysis. In
particular, we will show that the mean-field assumptions underlying
the TDT approach are valid for large critical nuclei $i\!\ge\!3$,
while for small critical nuclei $i\!=\!1,2$ second layer nucleation is
dominated by fluctuations.  Furthermore, we develop a novel rate
equation approach, which allows one to calculate the time-development
of cluster configurations on compact two-dimensional islands under
quite general conditions.

{F}rom the outset, one should distinguish between nucleation on top of
islands with compact shape as opposed to nucleation on islands with
strongly ramified shape. In the latter situation, diffusion of adatoms
on the islands becomes a rather complex phenomenon due to the confined
motion along branches of various
lengths.\cite{Havlin/Ben-Avraham:1987} We restrict our discussion to
nucleation on compact islands here. Intuitively, one would expect
second layer nucleation on ramified islands to be unlikely, so that
this restriction is of minor importance.  Moreover, it should be noted
that even for compact shapes, the island boundaries may have a fractal
or, more precisely, self-affine \cite{Barabasi/Stanley:1995} structure
(this is the case e.g.\ for Eden clusters \cite{Plischke/Racz:1984}).
Then the microscopic step edge barrier \cite{Feibelman:1998} can vary
strongly along the island boundary. In any case we will always
understand $\Delta E_{\rm S}$ as an effective barrier (see below and
ref.~\onlinecite{effes-comm}).

Parameters governing the second layer nucleation are the incoming atom
flux $F$, the jump rate $D/a^2$ of adatoms, the step edge barrier
$\Delta E_{\rm S}$, and various dissociation rates of unstable
clusters of size $s\!\le\!i$. If the bond energies of the unstable
clusters (i.e.\ of clusters of size $s\!\le\!i$) are negligibly small,
then the nucleation rate $\Omega(R)$ and critical radius $R_c$ depend
only on two dimensionless parameters. These are the ratio
\begin{equation}
\Gamma\equiv\frac{D}{Fa^4}
\label{gamma-eq}
\end{equation} 
(with $a$ being the lattice spacing in the substrate
plane) and the edge crossing probability 
\begin{equation}
\alpha\equiv\exp\left(-\frac{\Delta E_{\rm S}}{k_{\rm B}T}\right)\,.
\label{alpha-eq}
\end{equation} 
Based on a simplified stochastic description (see also
ref.~\onlinecite{Rottler/Maass:1999}) we will argue that for small
critical nuclei $i\!=\!1,2$ the mean number of atoms on top of the
island is smaller than one and the stable nucleus is formed due to
fluctuations. This gives rise to four scaling regimes in an
$\alpha-\Gamma$ diagram, where $R_c\sim \Gamma^\gamma\alpha^\mu$ with
different exponents $\gamma$ and $\mu$. For $i\!\ge\!3$ by contrast,
nucleation starts out from a situation with many atoms present on the
island. Under these circumstances, three different scaling regimes can
be identified, and two of them correspond to the ones predicted by the
TDT approach. By comparing $R_c$ with the mean distance $l$ between
islands on the substrate surface, the transition line separating rough
multilayer from smooth layer-by-layer growth is identified in the
$\alpha-\Gamma$ diagram. When the bond energies of unstable clusters
become appreciable, the corresponding dissociation rates enter the
problem as additional relevant parameters. It becomes difficult then
to separate scaling regimes in practice, and the simplified stochastic
description becomes of limited value. However, by employing the novel
rate equation approach it is still possible to determine $f_0(t)$ and
$R_c$ in a simple manner.

Moreover, we will discuss how to derive the fraction of $f(t)$ of
covered islands, when one relaxes the assumption that all island radii
evolve as the mean radius $R(t)$. In the time regime of almost
constant island density (``saturation regime'' preceeding island
coalescence\cite{Amar/Family:1996}) we can define an effective
``capture area'' for adatoms by the Voronoi cell for each
island.\cite{voronoi-comm} In order to calculate $f(t)$ for the
``multi-island model'' from $f_0(t)$, one needs to know the probability
distribution of islands with a certain size and capture area, when the
saturation regime is reached.

The paper is organized as follows. In Sec.~\ref{basic-sec} we give
a short review on basic concepts used in the description of
submonolayer growth and discuss in part \ref{thinfilm-subsec} important
quantities and equations underlying the physical processes involved.
In part \ref{TDT-subsec} we summarise the results of the TDT approach.
We then continue with a detailed description of the simulation
techniques in Sec.~\ref{sim-sec} and test the equivalence of the
multi-island and single-island model. 

In Sec.~\ref{simple-sec} a simplified stochastic description of second
layer nucleation is first presented in its general methodology and
subsequently employed to small and large critical nuclei. It is
instructive to observe the different physical conditions that lead to
the nucleation event in the two cases. After showing that the
treatment of metastable clusters is rather complicated within the
simplified description, we discuss in Sec.~\ref{general-sec} a general
approach for second layer nucleation on the basis of novel rate
equations.  Sec.~\ref{summary-sec} concludes the paper with a summary
and discussion of the most important results as well as an outlook to
further research.

\section{Basic quantities and concepts}
\label{basic-sec}

\subsection{Atomistic processes in thin film growth}
\label{thinfilm-subsec}
In MBE atoms are deposited on a substrate surface with a rate $Fa^2$
per unit cell. At very high temperatures the adatoms reevaporate but
under ordinary conditions this reevaporation can be neglected or
effectively taken into account by a reduced deposition rate.  Once an
adatom is deposited, it starts a thermally activated diffusive motion
with jump rate $D/a^2\propto \exp(-E_0/k_{\rm B}T)$.  Adatoms come
into contact as time progresses, and form islands that are held
together by some bonding energy. While unstable islands of small size
$s\le i$ dissociate, islands with size $s>i$ are stable (on all
relevant time scales of the experiment). An island of size $s=i$ is
called a critical nucleus.\cite{nucleus-comm}

Islands of larger size are formed by aggregation of adatoms (or small
mobile islands) to existing immobile islands, and by coalescence.
These processes lead to an island size distribution which becomes
broader with increasing time.  In the submonolayer regime, the most
important physical questions are: {\it (i)} How large is the density
$\rho_x(t)$ of stable islands on the substrate surface at time $t$?
{\it (ii)} What is the form of the distribution $\chi_s(s,t)$ of
island sizes $s$ at time $t$ (with $s$ being the number of atoms
forming the island)? {\it (iii)} What do the stable islands look like?
These questions have been extensively studied in the past, both by
experiment and by theory. We will briefly summarise those results,
which are relevant for the following analysis.

\begin{figure}[t]
\begin{center}\hspace*{-1cm}
\epsfig{file=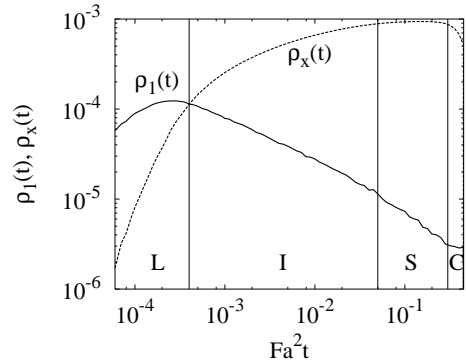,width=7cm}\end{center}
\vspace*{0.2cm}
\caption{The densities $\rho_1(t)$ and $\rho_x(t)$ of adatoms and
  stable islands as obtained in a kinetic Monte Carlo simulation
  modeling submonolayer growth (for an analogous figure see e.g.\ 
  ref.~\protect\onlinecite{Amar/Family:1996}). The different time
  regimes of low coverage L, intermediate coverage I, saturation S and
  coalescence C are indicated.}
\label{n1nx-fig}
\end{figure}

The typical behavior of $\rho_x(t)$ is depicted in Fig.~\ref{n1nx-fig}.
Also shown is the adatom density $\rho_1(t)$.  As suggested by Amar and
Family, \cite{Amar/Family:1996} one may distinguish between four
different time regimes: The low-coverage regime $L$ where $\rho_x(t)$
increases with $t$ and $\rho_x(t)\!<\!\rho_1(t)$, the intermediate coverage
regime I, where $\rho_x(t)$ increases with $t$ and $\rho_x(t)\!>\!\rho_1(t)$,
the saturation regime S (called aggregation 
%****************************************************************
\noindent
%****************************************************************
regime A in ref.~\onlinecite{Amar/Family:1996}), where $\rho_x(t)$
stays approximately constant, and the coalescence regime C, where
$\rho_x(t)$ strongly decreases due to coalescence of stable clusters.
At high coverages $\theta\equiv Fa^2t$, the monomer density
$\rho_1(t)$ becomes small (see Fig.~1), and the standard rate
equations for submonolayer growth \cite{Venables/etal:1984} predict
$\rho_x$ to evolve as
\begin{eqnarray}
\rho_x(t)&\propto&
  \left(\frac{D}{Fa^4}\right)^{-\frac{\scriptstyle i}{\scriptstyle i+2}}
  (Fa^2t)^{\frac{\scriptstyle 1}{\scriptstyle i+2}}\,
  e^{E_i/(i+2)k_{\rm B}T}\nonumber\\
&&\hspace*{-0.1cm}=\Gamma^{-\frac{\scriptstyle i}{\scriptstyle i+2}}\,
  \theta^{\frac{\scriptstyle 1}{\scriptstyle i+2}}\,e^{E_i/(i+2)k_{\rm B}T}\,,
\label{nx1-eq}
\end{eqnarray}
where $E_i$ is the bonding energy of the critical nucleus in its
preferred atomic configuration.

It should be noted that eq.~(\ref{nx1-eq}) is not valid in the
saturation regime S, where the densities of islands with subcritical
and critical size are very small (unless there are metastable
subcritical nuclei). In this regime almost all adatoms being deposited
attach to preexisting stable islands, so that $\rho_x(t)$ stays constant,
$\rho_x(t)=\rho_x$. Within the standard rate equation approach, this effect
may be accounted for by a proper dependence of the ``capture numbers''
on the adatom density $\rho_1(t)$ (for a detailed discussion of this
point in relation to experiments see
ref.~\onlinecite{Brune/etal:1999}). The scaling of $\rho_x$ with
$\Gamma$, however, is still correct in regime S,
\begin{equation}
\rho_x\propto\Gamma^{-\frac{\scriptstyle i}{\scriptstyle i+2}}\,.
\label{nx2-eq}
\end{equation}

One of the most detailed studies of the island size distribution has
been performed by Amar and Family \cite{Amar/Family:1996} based on the
scaling ansatz \cite{Tang/etal:1991,Bartelt/Evans:1992,Bales/Chrzan:1994}
\begin{equation}
\chi_s(s,t)=\frac{1}{\langle s(t)\rangle}\,
           f\Bigl(\frac{s}{\langle s(t)\rangle}\Bigr)\,.
\label{chi-eq}
\end{equation}
Here $\chi_s(s,t)=\rho_s(t)/\rho_{\rm tot}(t)$ is the probability that
an island has size $s$; ($\rho_s(t)$ is the density of islands with
size 
%****************************************************************
\begin{figure}[t]
\begin{center}\epsfig{file=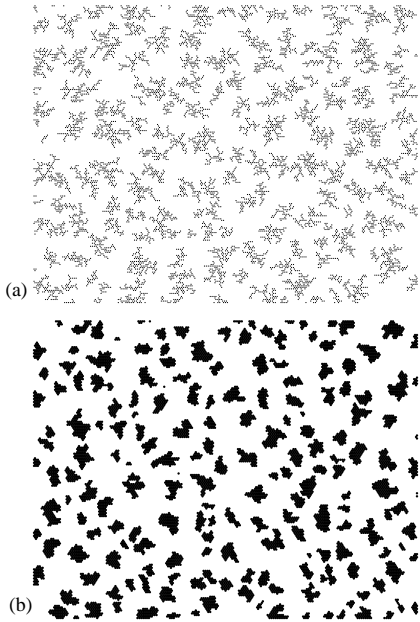,width=6cm}\end{center}
\vspace*{0.2cm}
\caption{Typical island morphologies as obtained in kinetic
  Monte Carlo simulations {\it (a)} for strictly irreversible
  attachment and {\it (b)} for irreversible attachment with local
  relaxation, in the absence of preferred growth directions.  For the
  simulation technique of the local relaxation process see
  Sec.~\ref{sim-sec} and Fig.~\ref{relax-fig}.}
\label{shapes-fig}
\end{figure}\noindent
%****************************************************************
$s$ and $\rho_{\rm tot}(t)$ is the total island density).
$\langle\ldots\rangle$ denotes an average over $s$ with respect to
$\chi_s(s,t)$. From (\ref{chi-eq}) follows, 
for $\langle
s^m\rangle\!<\!\infty$, $\langle s^m\rangle=\langle s\rangle^m
\int_0^\infty dx\, x^mf(x)$, and taking $m=0,1$ one obtains $\int dx
f(x)=\int dx\, x f(x)=1$. The mean island size $\langle s(t)\rangle$
is given by $\langle s(t)\rangle=\sum_{s=1}^\infty s
\rho_s(t)/\rho_{\rm tot}(t)=\theta/\rho_{\rm tot}(t)a^2$. In the
saturation regime S, in particular, where $\rho_{\rm tot}\simeq
\rho_x$, since the density of islands with subcritical and critical
size is small, it follows from eq.~(\ref{nx2-eq})
\begin{equation}
\langle s(t)\rangle\simeq\frac{\theta}{\rho_xa^2}\propto
\theta\,\Gamma^{\frac{\scriptstyle i}{\scriptstyle i+2}}\,.
\label{save-eq}
\end{equation}
The relation $\langle s(t)\rangle\simeq  theta/\rho_xa^2=Ft/\rho_x$ can also
be understood more directly, since the increase of $\langle
s(t)\rangle$ with $t$ is given by the flux times the mean capture area
$\rho_x^{-1}$ of adatoms.

The scaling function $f(u)$ was suggested to have the form
\cite{Amar/Family:1994}
\begin{equation}
f(u)=C_iu^i\exp(-i a_i u^{1/a_i})
\label{f(u)-eq}
\end{equation}
in regime S. This function has a maximum at $u\!=\!1$ and the two
conditions $\int dx f(x)=\int dx\, x f(x)=1$ determine the parameters
$C_i$ and $a_i$. Equations (\ref{chi-eq}-\ref{f(u)-eq}) have been
shown to give a fairly good approximation of some simulations and
experiments.

The problem of the island shapes is not yet well
understood,\cite{Brune:1998} but one may roughly answer the third
question posed above as follows. For strictly irreversible attachment,
where local relaxation of atoms due to fast edge diffusion is
suppressed, one obtains dendritic or random fractal structures.
Dendritic growth is preferred at low $T$ or small $F$, and a shape
transition from dendritic to random fractal structures has been found
e.g.\ for Ag/Pt(111) upon lowering the deposition
flux.\cite{Roeder/etal:1993} An example for a random fractal structure
obtained in a computer simulation is shown in Fig.~\ref{shapes-fig}a.
At high temperatures, edge diffusion becomes relevant, and polygonal
or ``irregular'' compact island morphologies develop (see
e.g.~ref.~\onlinecite{Michely/etal:1993}). An example for an irregular
structure is shown in Fig.~\ref{shapes-fig}b.

As mentioned in the Introduction, for the study of second layer
nucleation we will focus on compact island shapes.  Moreover, second
layer nucleation in the intermediate regime I is unlikely to occur,
since the island radii in this regime are typically smaller than the
critical radius $R_c$. We therefore consider the second layer
nucleation in the saturation regime S, where
eqs.~(\ref{nx2-eq}-\ref{f(u)-eq}) apply.

\subsection{TDT Approach}
\label{TDT-subsec}
In the TDT approach, \cite{Tersoff/etal:1994} one starts by calculating
the adatom density $\rho_1^{\scriptscriptstyle\rm st}$ on a circular
island with radius $R$ in the stationary state. The stationary
diffusion equation with the incoming atom flux acting as a source term
reads
\begin{equation}
D\left[\frac{\partial^2}{\partial r^2}+
\frac{1}{r}\frac{\partial}{\partial r}\right]
\rho_1^{\scriptscriptstyle\rm st}+F=0\,,
\label{diff-eq}
\end{equation}
and it is supplemented by the boundary conditions
($\alpha=\exp(-\Delta E_{\rm S}/k_{\rm B}T)$)
\begin{equation}
\frac{\partial \rho_1^{\scriptscriptstyle\rm st}}{\partial r}\Bigl|_{r=0}=0\,,
\hspace*{0.3cm}
-\frac{\partial \rho_1^{\scriptscriptstyle\rm st}}{\partial r}\Bigl|_{r=R}=
\frac{\alpha}{a}\, \rho_1^{\scriptscriptstyle\rm st}\Bigl|_{r=R}\,,
\label{bound-eq}
\end{equation}
where $a/\alpha$ is commonly referred to as the ``Schwoebel
length''.\cite{continuum-comm} The boundary conditions express the
fact that the current density $-D\partial \rho_1^{\scriptscriptstyle\rm
  st}/\partial r$ must vanish at the origin and that at the edge it is
given by the density $\rho_1$ times the ``velocity'' (rate times lattice
spacing) $(D\alpha/a^2)a$ to cross the step edge barrier. The
solution of eqs.~(\ref{diff-eq},{\ref{bound-eq}}) is
\begin{equation}
\rho_1^{\scriptscriptstyle\rm st}(r)\!=\!
\rho_1^{\scriptscriptstyle\rm st}(0)\!-\!\frac{Fr^2}{4D}\,,
\hspace*{0.5cm}
\rho_1^{\scriptscriptstyle\rm st}(0)\!=\!\frac{FR^2}{4D}
\left(1\!+\!\frac{2a}{\alpha R}\right)\,.
\label{rho-st-eq}
\end{equation}
According to standard rate equation theory \cite{Venables/etal:1984}
the local nucleation rate is proportional to
$D\rho_1^{i\!+\!1}$, so we obtain from eq.~(\ref{rho-st-eq})
\begin{eqnarray}
\Omega(R)&=&\kappa\frac{D}{a^2}\int_0^R\, 
\frac{2\pi r dr}{a^2}
[\rho_1^{\scriptscriptstyle\rm st}(r)a^2]^{i\!+\!1}\nonumber\\
&=&\frac{4\pi\kappa\Gamma^{-(i\!+\!1)}}{(i\!+\!2)\alpha^{2(i\!+\!2)}}
\frac{D}{a^2}
\left(\frac{\alpha R}{2a}\right)^{i\!+\!2}
\left[\left(1\!+\!\frac{\alpha
      R}{2a}\right)^{i\!+\!2}\!-\!1\right]\nonumber\\[0.15cm]
&\simeq&\left\{\begin{array}{ll}
\displaystyle4\pi\kappa\Gamma^{-(i\!+\!1)}
             \alpha^{-(i\!+\!1)}\frac{D}{a^2}
                   \left(\frac{R}{2a}\right)^{i\!+\!3}\,, 
           & \displaystyle\alpha\ll\frac{2a}{R}\\[0.5cm]
\displaystyle\frac{4\pi\kappa\Gamma^{-(i\!+\!1)}}{i\!+\!2}\frac{D}{a^2}
      \left(\frac{R}{2a}\right)^{2(i\!+\!2)}\,, 
           & \displaystyle\alpha\gg\frac{2a}{R}\end{array}\right.
\hspace*{-0.2cm}
\label{om-tdt-eq}
\end{eqnarray}
where $\kappa$ is a constant.

For a given time evolution of the island radius $R\!=\!R(t)$, one can
calculate the probability $f_0(t)$ for a stable nucleus to have formed
on top of the island up to time $t$ as follows: The increase
$f_0(t+\Delta t)\!-\!f_0(t)$ in a small time interval $\Delta t$ is equal
to the probability $[1-f_0(t)]$ that up to time $t$ no stable nucleus
has formed times the probability $\Omega(R(t))\Delta t$ that the
nucleation takes place in the time interval $[t,t\!+\!\Delta t]$.  Taking
the limit $\Delta t\to0$ and solving the corresponding differential
equation with the initial condition $f_0(0)\!=\!0$ yields
\begin{equation}
f_0(t)=1-\exp\left[-\int_0^t dt'\,\Omega(R(t'))\right]\,.
\label{f01-eq}
\end{equation}

For compact island growth during an MBE experiment, we have
$R(t)\!\sim\!\langle s(t)\rangle^{1/2}$ and thus from eq.~(\ref{save-eq})
\begin{equation}
\frac{R(t)}{a}=A(Fa^2t)^{1/2}\Gamma^{i/2(i\!+\!2)}
\label{R(t)-eq}
\end{equation}
with $A$ being some constant. Inserting this
growth law into (\ref{om-tdt-eq},\ref{f01-eq}) yields
\begin{eqnarray}
\label{om2-eq}
f_0(t)&=&1-\exp\left[-\frac{2\Gamma^{\scriptstyle-\frac{i}{i\!+\!2}}}
{A^2Fa^4}\int_0^{R(t)} dr\,r\Omega(r)\right]\\[0.3cm]
&&\hspace*{-1cm}\simeq\left\{\begin{array}{ll}
\displaystyle  1-\exp\left[-C_{\scriptscriptstyle <}
         \Gamma^{\scriptstyle-\frac{i(i\!+\!3)}{i\!+\!2}}
              \alpha^{-(i\!+\!1)}
                        \Bigl(\frac{R}{a}\Bigr)^{i\!+\!5}\right]\,,
    & \displaystyle \alpha\ll \frac{2a}{R(t)}\\[0.6cm]
\displaystyle 1-\exp\left[-C_{\scriptscriptstyle >}
        \Gamma^{\scriptstyle-\frac{i(i\!+\!3)}{i\!+\!2}}
      \Bigl(\frac{R}{a}\Bigr)^{2(i\!+\!3)}\right]\,,
    & \displaystyle \alpha\gg \frac{2a}{R(t)}\end{array}\right.\nonumber
\end{eqnarray}
where $C_{\scriptscriptstyle >}\!\equiv\!(2^{-2(i\!+\!1)}\pi\kappa
A^{-2})/[(i+2)(i+3)]$ and $C_{\scriptscriptstyle <}\!\equiv\!
2^{-i}\pi\kappa A^{-2}$. In going from the first to the second line in
(\ref{om2-eq}) we have used that the integral over $r$ is dominated by
the upper integration bound $R(t)$ (for $R(t)/a\gg1$). It follows
that the critical radius scales as
\begin{equation}
R_c\sim\Gamma^\gamma\alpha^\mu\,,
\label{gamma-mu-eq}
\end{equation}
where
\begin{equation}
\gamma=\left\{\begin{array}{l@{\hspace{0.3cm}}l}
\displaystyle\frac{i(i\!+\!3)}{(i\!+\!2)(i\!+\!5)}\,, & 
\displaystyle\alpha\ll \Gamma^{i/[2(i\!+\!2)]}\\[0.4cm]
\displaystyle\frac{i}{2(i\!+\!2)}\,, & 
\displaystyle\alpha\gg \Gamma^{i/[2(i\!+\!2)]}\end{array}\right.
\label{gamma-tdt-eq}
\end{equation}
and
\begin{equation}
\mu=\left\{\begin{array}{l@{\hspace{0.3cm}}l}
\displaystyle\frac{(i\!+\!1)}{(i\!+\!5)}\,, & 
\displaystyle\alpha\ll \Gamma^{i/[2(i\!+\!2)]}\\[0.4cm]
\displaystyle0\,, & 
\displaystyle\alpha\gg \Gamma^{i/[2(i\!+\!2)]}\end{array}\right.
\label{mu-tdt-eq}
\end{equation}
Equations (\ref{gamma-mu-eq}-\ref{mu-tdt-eq}) predict that for large
step edge barriers, $R_c$ depends strongly on $\Delta E_{\rm S}$,
$R_c\sim\exp[-(i+1)\Delta E_{\rm S}/(i+5)k_{\rm B}T]$, while for small
barriers, $R_c$ becomes independent of $\Delta E_{\rm S}$. For $i=1$ in
particular, one finds $R_c\sim\Gamma^{2/9}\alpha^{1/3}$ for
$\alpha\ll\Gamma^{1/6}$ and $R_c\sim\Gamma^{1/6}$ for
$\alpha\gg\Gamma^{1/6}$.

\section{Kinetic Monte Carlo Simulations}
\label{sim-sec}
Kinetic Monte Carlo simulations are a well-established technique for
modeling MBE
experiments.\cite{Bartelt/Evans:1999,Bott/etal:1996,Amar/etal:1994,Liu/etal:1993}
In our investigation of second layer nucleation, we adopt a simulation
scheme similar to previous, successful models of surface growth
kinetics.\cite{Amar/etal:1994,Bales/Chrzan:1994} We chose a substrate
with fcc(111) symmetry, since surfaces of that kind are often studied
in metal epitaxy, and commonly exhibit high $\Delta E_{\rm S}$. In a
full simulation scheme of the growth kinetics (multi-island model), we
include all processes of evaporation, diffusion and aggregation
occurring in the MBE experiment. By analyzing the set of islands of
various size on the substrate, we determine the fraction $f(t)$ of
covered islands at time $t$. On the other hand, we consider, as in the
TDT approach, only one island with the mean radius $R(t)$ evolving
deterministically in time. The fraction $f_0(t)$ of covered islands in
this single-island model is then determined by calculating the
probability for second layer nucleation up to time $t$ from a large
set of independent simulations.  By examining both models we are able
to quantify the influence of the cluster size distribution under
generic growth conditions.

\subsection{Multi-island model}
\label{multi-subsec}
Atoms are randomly deposited with a rate $Fa^2$ per unit cell onto a
triangular lattice. After instantaneously relaxing to a position,
where they are supported by three nearest neighbors in the layer below
(``downward funneling''), the atoms change their position by
performing thermally activated jumps to a vacant nearest neighbor site
in the same layer with a rate $D/6a^2$. Only one atom is allowed to
occupy a given lattice site.  We first consider a ``non-interacting
particle model'' where all binding energies of subcritical clusters of
size $s\!\le\!i$ are neglected. This means that $(i\!+\!1)$ adatoms
have to encounter each other on nearest neighbor sites in order to
form a stable nucleus. Within the single-island model we later will
also consider finite binding energies of metastable clusters, which
causes various dissociation rates to enter the problem as additional
parameters.

Once a stable cluster of size $s\!>\!i$ has formed, adatoms can attach
to it. Compact island morphologies are known to emerge if a fast
diffusion process is present along island edges.  Here we model this
process similar as in earlier approaches (see e.g.\ 
refs.~\onlinecite{Amar/Family:1996,Bartelt/Evans:1995}) by including a
local relaxation mechanism. In this method an atom being in
%****************************************************************
\begin{figure}[t]\hspace*{1.5cm}
\begin{center}\epsfig{file=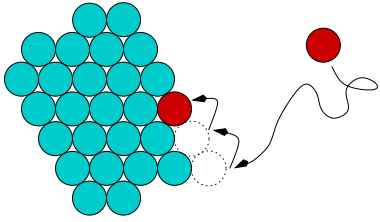,width=5cm}\end{center}
\vspace*{0.2cm}
\caption{Illustration of the local relaxation process when an adatom
  arrives at an island edge.}
\label{relax-fig}
\end{figure}\noindent
%****************************************************************
contact with at least one nearest neighbor after a jump, is immediately
transferred to a nearest neighbor site, if it can increase its coordination
number. This procedure is repeated until the atom can no longer increase its
local coordination (see Fig.~\ref{relax-fig}).

Interlayer diffusion of atoms deposited onto islands is hindered by
the Ehrlich-Schwoebel barrier $\Delta E_{\rm S}$, which reduces the
jump rate $D/6a^2$ by the edge crossing probability
$\alpha=\exp(-\Delta E_{\rm S}/k_{\rm B}T)$.\cite{alpha-comm} For
computational convenience, we model the crossing by a two-step process
in the simulation: First, when an atom passes the boundary, it remains
in the same layer but moves to a place, where it is supported by only
two atoms underneath.  Then the atom immediately drops down to the
layer below and moves to the nearest ``stable'' site according to the
local relaxation mechanism introduced above (for similar simulations
including $\Delta E_{\rm S}$, see,
e.g.~\onlinecite{Smilauer/Harris:1995,Ratsch/etal:1994}). We do not
distinguish between crossing of A and B steps and have not attempted
to model any more realistic scenarios, as e.g.\ collective
rearrangements of atoms including exchange
processes.\cite{Wang/Ehrlich:1991,Stumpf/Scheffler:1994,Jacobsen/etal:1995}
This is well justified as long as one is interested in the influence
of an effective Schwoebel barrier.\cite{effes-comm}

\begin{figure}[t]
\begin{center}\epsfig{file=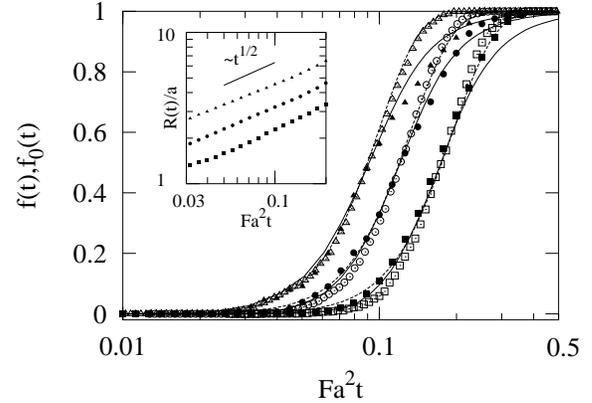,width=8cm}\end{center}
\vspace*{-0.2cm}
\caption{The fractions $f(t)$ (full symbols) and
  $f_0(t/1.21)$ (open symbols) of covered island as a function of the
  total coverage $Fa^2t$ for $i\!=\!1$, $\alpha\!=\!10^{-5}$, and
  three different $\Gamma\!=\!10^5$ ($\scriptscriptstyle
  \blacksquare$, $\scriptscriptstyle\square$), $10^6$ ($\bullet$,
  $\circ$), and $10^7$ ($\blacktriangle$, $\triangle$).  The dashed
  lines fitting $f_0(t/1.21)$ were calculated from the theoretical
  predictions for the second layer nucleation rate $\Omega(R(t))$
  (eq.~(\ref{om-fl-eq})). The solid lines fitting $f(t/1.21)$ were
  calculated according to eq.~(\ref{f-f0--simple-eq}) with
  $\chi_\sigma(\sigma)\!\sim\!\sigma^2$ (see text). The inset displays
  the time-dependence of the mean island radius
  $R(t)\!\equiv\!(\langle s(t)\rangle/\pi)^{1/2}a$ (see
  eqs.~(\ref{save-eq},\ref{R(t)-eq})).}
\label{f-t-fig}
\end{figure}

In the following, we focus on the case $i\!=\!1$ first. Typical film
morphologies resulting from the simulations have been shown in
Fig.~\ref{shapes-fig}b. Note that the boundaries of the
islands are still
rough despite the local relaxation mechanism. The fraction $f(t)$ of covered
islands as a function of the total coverage $Fa^2t$ is shown in
Fig.~\ref{f-t-fig} for some representative parameters (full symbols). As
expected, $f(t)$ first is close to zero, then increases strongly in some time
interval around a ``critical time'' $t_c$, and finally saturates
at one. In the inset of
Fig.~\ref{f-t-fig} we show the dependence of the mean island radius
$R(t)\!\equiv\!(\langle s(t)\rangle/\pi)^{1/2}a$ on $Fa^2t$ during the
evaporation. In agreement with eq.~(\ref{R(t)-eq}), we find
$R(t)=A(Fa^2t)^{1/2}\Gamma^{1/6}$ with $A\!\cong\!0.78$.

\begin{figure}[t]
\begin{center}\epsfig{file=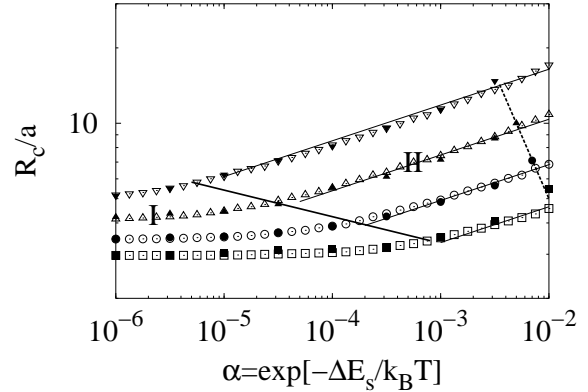,width=8cm}\end{center}
\vspace*{-0.2cm}
\caption{Dependence of the critical island radius $R_c$ on $\alpha$
  for $i\!=\!1$ and various $\Gamma=10^5$
  ($\scriptscriptstyle\blacksquare$, $\scriptscriptstyle\square$),
  $10^6$ ($\bullet$, $\circ$), $10^7$ ($\blacktriangle$, $\triangle$),
  and $10^8$ ($\blacktriangledown$, $\triangledown$). Full symbols
  refer to the results from the multi-island model, while open symbols
  refer to the results obtained from the single-island model
  ($R_c=1.1R_c'$). The dashed line marks the onset of layer-by-layer
  growth, and the solid line with negative slope marks the crossover
  between regimes I and II; the solid lines fitting the data in regime
  II have slope 1/7.}
\label{rc-I-II-fig}
\end{figure}

To be specific, let us define the critical time $t_c$ via the
condition $f(t_c)\!=\!1/2$, and the corresponding critical island
radius $R_c$ by $R_c\!=\!R(t_c)$,
\begin{equation}
f(t_c)=1/2\,,\hspace*{0.15cm}
R_c\equiv R(t_c)=A(Fa^2t_c)^{1/2}\Gamma^{i/2(i\!+\!2)}a\,.
\label{tcrc-eq}
\end{equation}\noindent
Plots of $R_c$ as a function of $\alpha$ for various fixed $\Gamma$ are shown
in Fig.~\ref{rc-I-II-fig} (full symbols). With increasing step edge barrier,
i.e.\ decreasing $\alpha$, adatoms on average remain longer
%****************************************************************
\vspace*{-0.8cm}
%****************************************************************
\noindent
%****************************************************************
on an island and nucleation of stable dimers occurs at smaller island radii.
Accordingly, $R_c$ decreases with decreasing $\alpha$ (see ``regime II'' in
the figure). For very small $\alpha$, however, the step edge barrier is
practically never surmounted and thus is in effect infinitely high. Therefore,
$R_c$ becomes independent of $\alpha$ (``regime I'' in
Fig.~\ref{rc-I-II-fig}).  The crossover between the two regimes is marked by
the thick solid line.  The full symbols in Fig.~\ref{rc-I-II-fig} terminate at
the dashed line $\alpha_\star(\Gamma)$, which marks the onset of island
coalescence. For $\alpha\!>\!\alpha_\star(\Gamma)$, islands in the first layer
merge before second layer nucleation takes place and $R_c$ can no longer
determined from the multi-island model. It is important to note that the
dependence of $R_c$ on $\alpha$ is much weaker than predicted by the TDT
approach: The solid lines in regime II have slope 1/7 corresponding to a power
law $R_c\sim\alpha^{1/7}$ rather than $R_c\sim\alpha^{1/3}$ as predicted by
eqs.~(\ref{gamma-mu-eq}, \ref{mu-tdt-eq}). Moreover, regime I does not occur
in the TDT approach.

\subsection{Single-island model}
\label{single-subsec}
Second layer nucleation can also be addressed in a simpler model,
which does not attempt to describe the entire growth dynamics, but
focuses on the decisive factors that determine nucleation in the
presence of the step-edge barrier. In this model the complicated
nucleation and diffusion-mediated growth of the two-dimensional
islands, on which the second layer nucleation takes place, is replaced
by letting the radius of {\it one} circular island expand
deterministically in time as
$R(t)/a\!=\!A(Fa^2t)^{1/2}\Gamma^{i/2(i\!+\!2)}$, where $A$ is taken
from the full simulation of the multi-island model.

The island is embedded in a substrate area large enough to accommodate
the island at all relevant times. Deposition and diffusion of adatoms
take place in the same manner as in the multi-island model. Atoms
inside the island boundary can escape by overcoming the step edge
barrier. Those atoms that have surmounted the barrier or that have been
deposited outside the island boundary are removed from the lattice.
Thus the single-island model considers the deposition of random
walkers within a time-dependent, circular boundary that is partially
reflecting. Due to its greater simplicity, it allows for more specific
analysis with a larger parameter space (there is no restriction due to
coalescence of distinct islands).

In the non-interacting particle model, the ``critical event'' is to find
$(i\!+\!1)$ atoms on neighboring lattice sites.  Analogous to the multi-island
model we can define the fraction $f_0(t)$ of covered islands up to time $t$.
The fraction now refers to a set of islands obtained in independent simulation
runs. All islands in these runs grow with the same deterministic growth law.
Results for $f_0(t)$ are shown in Fig.~\ref{rc-I-II-fig} (open symbols) for
the same parameters as in the multi-island model. Good agreement with $f(t)$
is achieved for small times (corresponding to $f(t)\!\lesssim\!1/2$), when the
time in the single-island model is rescaled by a constant factor, i.e.\ 
$f(t)\!\simeq\!f_0(t')$ with $t'\!=\!t/1.21$. The factor is a consequence of
the idealized circular island perimeter in the single-island model. In the
multi-island model by contrast, the islands are far from being perfectly
circular (see Fig.~\ref{shapes-fig}). They have rougher edges with more
boundary sites, which causes adatoms to escape the islands more easily and
second layer nucleation to occur at later times $t\!\cong\!1.21t'$.

At larger times (corresponding to $f(t)\!\gtrsim\!1/2$), however,
$f(t)$ deviates from $f_0(t')$ and these deviations become more
pronounced for larger $\Gamma$. The reason for this discrepancy is the
presence of islands with size much smaller than $\langle s(t)\rangle$
in the multi-island model. Nucleation of stable clusters on top of
these islands occurs at a later time, which causes $f(t)$ to be
smaller than $f_0(t')\!=\!f_0(t/1.21)$ at large $t$.  In fact, we will
show in Sec.~\ref{equiv-subsec} that this effect can be accounted for
by considering the probability distribution of islands with a certain
size and capture area. When $R_c$ approaches the mean distance $l$,
coalescences of larger islands also lead to modifications of $f(t)$
for $t\gtrsim t_c$.

The critical radius $R_c'$ in the single-island model can be defined as
in the many island model by $R_c'\!=\!R(t_c')$, where
$f_0(t_c')\!=\!1/2$. Due to the fact that $t_c\!=\!1.21t_c'$ we expect
$R_c\!=\!1.21^{1/2}R_c'\!=\!1.1R_c'$.  Results for $1.1R_c'$ as a
function of $\alpha$ are shown in Fig.~\ref{rc-I-II-fig} (open
symbols) for the same parameters as in the multi-island model (full
symbols). As can be seen from the figure, there is almost perfect
agreement between both data sets. Moreover, the data for $R_c'$ can be
obtained also beyond the dashed line marking the onset of
layer-by-layer growth. Let us also note that, as long as one is
interested only in $R_c'$ (or $R_c=1.1R_c'$), one may obtain it even
more simply in the single-island model (without calculating $f_0(t)$) by
determining the average radius of the island at the time of the
nucleation event,
\begin{equation}
R_c'=R(t_c')\simeq\int_0^\infty dt\,
\frac{df_0(t)}{dt}\,R(t)\,.
\label{rcave-eq}
\end{equation}
Note that $df_0(t)/dt$ is the probability density of the second layer
nucleation times and that the average of $R(t)$ with respect to $df_0(t)/dt$
is approximately equal to $R(t_c')$, since $df_0(t)/dt$ is sharply peaked
around $t_c'$.

\subsection{Equivalence of the single-island and the multi-island model}
\label{equiv-subsec}
In order to determine $f(t)$ from $f_0(t)$ we define by
$\psi(s,\sigma,t)\,ds\,d\sigma$ the probability for an island to have
a size in the interval $[s,s\!+\!ds]$ and a capture area in the
interval $[\sigma,\sigma\!+\!d\sigma]$ at time $t$, where the capture
area is given by the Voronoi cell associated with an
island.\cite{voronoi-comm}

Let us consider $f_0(t)$ to be a functional of the growth law $R(t)$
only, as it is the case, for example, when one approximates the second
layer nucleation by a Poisson process with a time dependent nucleation
rate $\Omega(R(t))$. Then $f_0(t)=G_0[R(t)]=1-\exp[-\int_0^t
dt'\,\Omega(R(t'))]$ (see eq.~\ref{f01-eq}).  In the saturation regime
the growth law for an island can be written as $\pi
R^2(t)=s_\times+F\sigma(t-t_\times)$, where $t_\times$ is the time
when the saturation is reached (see Fig.~\ref{n1nx-fig}) and
$s_\times$ is the island size at that time. (We restrict ourselves to
film-morphologies far from coalescence here, so that $\sigma$ can be
regarded as time-independent.)  With the specified growth law, the
functional $G_0[R(t)]$ can be expressed by a function
$g_0=g_0(t;s_\times,\sigma,t_\times)$, and $f(t)$ is calculated via
\begin{equation}
f(t)=\int_0^\infty\hspace*{-0.15cm} ds_\times 
\int_0^\infty \hspace*{-0.15cm} d\sigma\,\psi(s_\times,\sigma,t_\times)\,\,
g_0(t;s_\times,\sigma,t_\times)\,.
\label{f-f0-eq}
\end{equation}
A detailed investigation of the probability distribution $\psi(s,\sigma,t)$ is
certainly of interest but beyond the scope of the present work. A simple idea
would be to neglect correlations between the stochastic variables $s$ and
$\sigma$, $\psi(s,\sigma,t)\simeq \chi_s(s,t)\chi_\sigma(\sigma,t)$, and to
use previously derived scaling forms for the island size distribution
$\chi_s(s,t)$ (see e.g.\ 
refs.~\onlinecite{Bartelt/Evans:1999,Amar/Family:1996}) and the capture area
distribution $\chi_\sigma(\sigma,t)$ (see e.g.\ 
refs.~\onlinecite{Bartelt/Evans:1992,Stroscio/Pierce:1994}).

Here we will follow a simpler approach. Since for typical situations we find
both $s_\times$ and $t_\times$ to be significantly smaller than $s_c\!=\!\pi
R_c^2$ and $t_c$, respectively, we use the growth law $s(t)\!=\!F\sigma t$ for
an island with capture area $\sigma$ in the full simulation. The on-top
nucleation probabilities $\tilde g_0(t;\sigma)$ for islands exhibiting
different capture areas can then be related by a rescaling of time, i.e.\ 
$\tilde g_0(t;\sigma_1)= \tilde g_0(\sigma_1t/\sigma_2;\sigma_2)$. Moreover,
since for film morphologies far from coalescence
($\alpha\ll\alpha_\star(\Gamma)$), $\chi(\sigma,t)$ is approximately
independent of time, we have $f_0(t)\!=\!\tilde g_0(t;\bar\sigma)$, where
$\bar\sigma\!=\!\int d\sigma\chi_\sigma(\sigma,t)\sigma\!\simeq\!\rho_x^{-1}$.
Hence,
\begin{equation}
f(t)\!=\!\int_0^\infty\hspace*{-0.2cm}d\sigma\,
\chi_\sigma(\sigma)\,\tilde g_0(t;\sigma)=
\int_0^\infty\hspace*{-0.2cm} d\sigma\,\chi_\sigma(\sigma)\,
f_0\Bigl(\frac{\bar\sigma}{\sigma}t\Bigr)\,.
\label{f-f0--simple-eq}
\end{equation}
In this simplified eq.~(\ref{f-f0--simple-eq}) knowledge of the
nucleation rate $\Omega(R)$ is not necessary and $f(t)$ can be directly
obtained from $f_0(t)$ when $\chi_\sigma(\sigma)$ is known.

Writing $\chi_\sigma(\sigma)\!=\!\bar\sigma^{-1}h(\sigma/\bar\sigma)$,
where $\int dx\,h(x)\!=\!\int dx h(x) x\!=\!1$, the transformation
(\ref{f-f0--simple-eq}) becomes $f(t)\!=\!\int_0^\infty dx
h(x)f_0(xt)$. For a random distribution of point islands, we would have
$h(x)\!=\!\exp(-x)$. However, since there is a depletion zone of
adatoms near an island, the probability for other islands to nucleate
in an area close to an existing one is reduced and not exponential.
For an isolated island, dimensional analysis predicts the extension
$\xi$ of the depletion zone to be of order $(D/F)^{1/4}$. By
comparing $\xi$ with the mean distance $l\sim \rho_x^{-1/2}\sim
\Gamma^{i/2(i\!+\!2)}$ between islands, we expect that $h(x)$ exhibits
a large $x$ regime with $h(x)\sim\exp(-x)$ only for $i\!>\!2$. For
$i\!=\!1$ we thus are satisfied with a simple power law ansatz
$h(x)\!=\!Cx^{\phi}$ for $x\le x_{\star}$, where $C$ and $x_\star$
follow from the two conditions imposed on $h(x)$, and $\phi$ is a
fitting parameter.

To test this ansatz we take $f_0(t)$ for $\Gamma\!=\!10^7$ from
Fig.~\ref{f-t-fig} (open symbols or dotted lines) and compare $f(t)$
as calculated from eq.~(\ref{f-f0--simple-eq}) (solid lines in
Fig.~\ref{f-t-fig}) with the corresponding $f(t)$ as obtained in the
simulation (full symbols in Fig.~\ref{f-t-fig}).  As can be seen from
Fig.~\ref{f-t-fig}, for $\Gamma\!=\!10^6$ and $\Gamma\!=\!10^7$ a
fairly good account of the differences between $f_0(t)$ and $f(t)$ can
be obtained by choosing $\phi\!=\!2$.  However, for $\Gamma\!=\!10^5$
the theoretical curve underestimates the fraction of covered islands
at large times (where $f(t)\gtrsim1/2$). Better agreement between
theory and simulation can only be obtained if one would allow $\phi$
to depend on $\Gamma$. Alternatively, we have tried an ansatz for
$h(x)$ similar to that used by Amar and Family for the scaling
function characterizing the island size distribution (see
eq.~(\ref{f(u)-eq})). This ansatz yields comparable results, but is
also not successful in accounting for the changes with $\Gamma$.  We
finally have to note, that at the time when submitting the paper, a
more detailed theoretical account for $\psi(s,\sigma,t)$ was
published.\cite{Mulheran/Robbie:2000} The use of this finding in
eq.(\ref{f-f0-eq}) and the comparison of the resulting $f(t)$ with Monte
Carlo data will be presented elsewhere.

Having shown that the single-island and multi-island models are
essentially equivalent, except for differences between $f(t)$ from
$f_0(t)$ for large times that can be attributed to the island size
distribution, we will focus on the single-island model in the
remaining part of the paper.

\section{Second layer nucleation in simple situations}
\label{simple-sec}
In this Section we develop a stochastic description of the nucleation
process based on the scaling approach for second layer nucleation
presented in ref.~\onlinecite{Rottler/Maass:1999} (see also
ref.~\onlinecite{Krug/etal:2000}). The procedure focuses on the
non-interacting particle model, although formally it is possible to
extend scaling concepts also to situations, where the lifetimes of
unstable clusters become important. This was shown by Krug {\it et
  al.}\cite{Krug/etal:2000} and is discussed in a more general context
in Sec.~\ref{simple-meta-subsec}.  The treatment of the
non-interacting particle model outlined in
Sec.~\ref{simple-proc-subsec} already captures the salient features of
the problem in terms of lifetimes, occupation probabilities and
encounter rates.  We will show that there exist two possible
mechanisms for the formation of a stable cluster: In the first case,
there is typically no atom on top of the island and a stable cluster
is formed due to fluctuations, in which by chance $i\!+\!1$ atoms are
present on the island. In the second case by contrast, there are on
average more than $i\!+\!1$ atoms on top of the island during the
formation of a stable cluster so that the nucleation process can be
described in a mean-field type manner.

It turns out that the fluctuation-dominated case takes place for
$i\!\le\!2$, while the mean-field situation occurs for $i\!\ge\!3$.
The TDT approach corresponds to the mean-field case with the notable
supplement that for very large step edge barriers one should deal with
the time-dependent adatom density $\rho_1(r,t)$ (solution of
eqs.~(\ref{diff-eq},\ref{bound-eq})) to calculate the nucleation rate
$\Omega(R)$ from eq.(\ref{om-tdt-eq}).  In the language of critical
phenomena, one may regard $i\!=\!2$ as the upper critical size of the
critical nucleus above which mean-field theory becomes applicable. We
have to note that the existence of this upper critical size was not
perceived by us in ref.~\onlinecite{Rottler/Maass:1999}, and
accordingly, the extension of the scaling arguments for the
fluctuation-dominated situation to $i\!=\!3$ was not allowed.

In the stochastic formulation presented below we will develop many of
the necessary ingredients for the general treatment of second layer
nucleation in the next Sec.~\ref{general-sec}. Moreover, it is discussed
under which conditions mean-field type expressions $\propto
D\rho_1^{i+1}$ for local nucleation rates can be used.

\subsection{General procedure}
\label{simple-proc-subsec}
In order to determine a second layer nucleation rate $\Omega(R)$ we
start by considering a time interval $\Delta t(R)$, during which
$R(t)$ does not change significantly.  For example, for the generic
growth law (\ref{R(t)-eq}) we may require $\Delta t(R)$ to correspond
to a 10\% change of $R$, which would give $\Delta
t(R)=0.21(R/a)^2/[A^2Fa^2\Gamma^{i/(i\!+\!2)}]$, i.e.\ 
\begin{equation}
\Delta t(R)\sim F^{-1}\Gamma^{-i/(i\!+\!2)}R^2
\label{deltat-eq}
\end{equation} 
The nucleation rate $\Omega(R)$ is the mean number
$n_{\scriptscriptstyle\rm nuc}(R)$ of nucleation events in time
$\Delta t(R)$ divided by $\Delta t(R)$,
\begin{equation}
\Omega(R)=\frac{n_{\scriptscriptstyle\rm nuc}(R)}{\Delta t(R)}\,.
\label{om-nnuc-eq}
\end{equation} 
A nucleation event occurs, if $i\!+\!1$ atoms encounter each other on nearest
neighboring sites.

For an island with radius $R$ and infinite step edge barrier
($\alpha=0$), and in total $n$ single atoms on top of it, let us
approximate the encounter dynamics by a Poisson process, where
$\omega_n(R)$ denotes the encounter rate of exactly $i\!+\!1$ atoms.
Within the Poisson approximation this rate can be precisely defined
as the inverse average time for $i\!+\!1$ atoms to encounter each
other for the first time, when initially $n$ atoms are randomly
distributed on top of the island. A simple scaling argument yields
\begin{equation}
\omega_n(R)=\kappa_{\rm e}\,\left[\prod_{k=0}^i
(n\!-\!k)\right]\,\frac{D}{a^2}\,
\left(\frac{a^2}{\pi R^2}\right)^{i\!+\!1}\,\frac{\pi R^2}{a^2}\,,
\label{omn-eq}
\end{equation}
where $\kappa_{\rm e}$ is a constant. The term $(a^2/\pi
R^2)^{i\!+\!1}$ is proportional to the probability to find $i\!+\!1$
atoms on nearest neighbor sites, and the factor $(\pi R^2/a^2)$ takes
into account that the encounter can occur everywhere on the island.
The combinatorial factor $\prod_{k=0}^i(n\!-\!k)$ is slightly more
subtle. At first sight, one may think that one should include the
number ${n\choose i\!+\!1}$ of possibilities to choose any $i\!+\!1$
atoms out of the $n$ atoms but this is not correct, since the
accumulation of $i\!+\!1$ atoms does not happen ``in parallel'' at a
certain instant of time but in order: First a dimer forms out of $n$
single atoms (combinatorial factor $n(n\!-\!1)/2$) and then some of
the remaining $n\!-\!k$ atoms ($k\!=\!2,3,\ldots,i$) have to attach
one after another to an intermediate cluster of size $k$ before this
cluster dissociates (the intermediate cluster is assumed to be much
less mobile than single adatoms).  The sequential attachment process
yields an additional combinatorial factor $\prod_{k=2}^i(n\!-\!k)$.
Clearly, the scaling argument gives only a 
%****************************************************************
\begin{figure}[t]
\begin{center}\epsfig{file=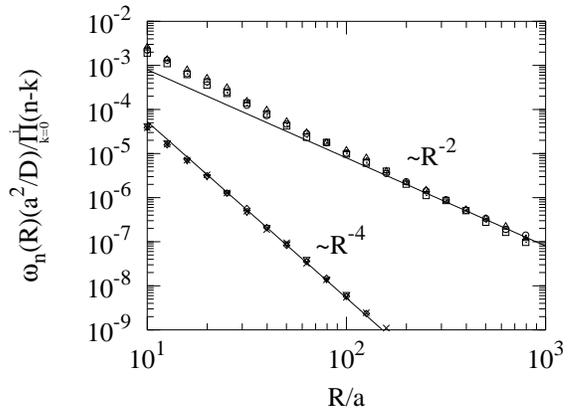,width=8cm}\end{center}
\vspace*{-0.4cm}
\caption{Scaled rate $\omega_n(R)(a^2/D)/\prod_{k=0}^i(n\!-\!k)$ 
of the encounter of $i\!+\!1$ atoms
out of $n$ atoms on an island with radius $R$ and infinite step edge barrier.
The upper curve with slope (-2) refers to $i\!=\!1$ and $n\!=\!2$
($\scriptscriptstyle\square$), 3 ($\circ$), and 4 ($\triangle$),
and the lower curve with slope (-4) refers to $i\!=\!2$ and $n\!=\!3$
($\triangledown$), 4 ($\lozenge$), and 5 ($\times$).}
\label{omn-fig}
\end{figure}\vspace*{-0.4cm}\noindent
%****************************************************************
rough approximation for
$\omega_n(R)$ and a more refined treatment justifying
eq.~(\ref{omn-eq}) is presented in Appendix~\ref{omn-app}.

Determination of $\omega_n(R)$ for $i\!=\!1,2$ and various $n$ in our
simulations confirms the behavior predicted by eq.~(\ref{omn-eq}), see
Fig.~\ref{omn-fig}. For $i\!=\!1$ the scaling law is only valid for
large $R\!\gtrsim\!100a$, because at smaller $R$, two atoms typically
encounter each other before the delta-functions characterizing the
initial occupancy smear out to a uniform distribution (for larger $i$
this effect becomes less important). Moreover, we find $\kappa_{\rm
  e}\!\cong\!0.087$ for $i\!=\!1$ and $\kappa_{\rm e}\!\cong\!0.53$
for $i\!=\!1$, i.e.\ the coefficient $\kappa_{\rm e}$ is constant for
fixed $i$, but changes strongly with $i$. This dependence is expected,
since we neglected the memory effect that, when $n$ atoms,
$2\!\le\!n\!\le\!i$, are already close to each other, they keep close
together for a while so that the encounter of $i\!+\!1$ atoms during
this intermediate time becomes more likely. This memory effect is not
included in the treatment in Appendix~\ref{omn-app}, where after each
``dissociation'' of an unstable cluster of size $k\!\le\!i$ a
configuration is assumed to emerge, where a cluster of size $k\!-\!1$
is left and the remaining $n\!-\!k$ atoms are assumed to be randomly
distributed.  Accordingly, $\kappa_{\rm e}$ should increase with
increasing $i$ as it is the case.

\begin{figure}[b]
\begin{center}\epsfig{file=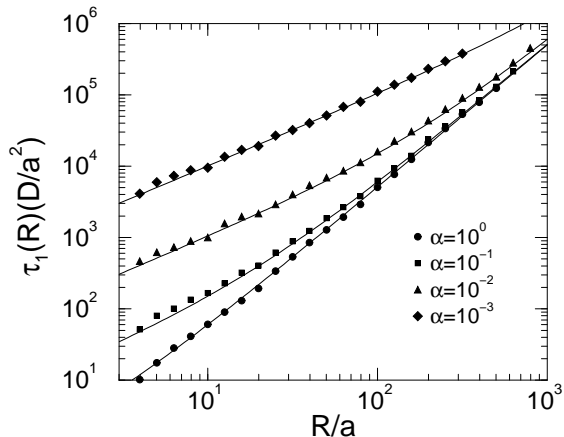,width=7.3cm}\end{center}
\vspace*{-0.5cm}
\caption{Lifetime $\tau_1(R)$ of a single atom on an
  island with radius $R$ in units of $a^2/D$ for various edge
  crossing probabilities $\alpha\!=\!\exp(-\Delta E_{\rm S}/k_{\rm
    B}T)$. The solid lines are drawn according to eq.~(\ref{taun-eq})
  with $\kappa_1\!=\!1$ and $\kappa_2\!=\!1/2$.}
\label{taun-fig}
\end{figure}

Equation~(\ref{omn-eq}) has been derived for an infinite step edge
barrier.  For finite step edge barriers, we have to take into account
that a state corresponding to an island with $n$ atoms on top of it
has a finite lifetime $\tau_n(R)$ only.  This lifetime is defined by
the average time required for the {\it first} of the $n$ atoms to
escape from the island (if any encounter processes are neglected). To
a good approximation, $\tau_n(R)$ is the $n$th fraction of the
lifetime $\tau_1(R)$ of a single atom,
$\tau_n(R)\!\simeq\!\tau_1(R)/n$ (this approximation would become
exact, if the escape were a simple Poisson process). In the limit of
large $\alpha$, $\tau_1(R)$ is proportional to the characteristic time
$R^2/D$ for an atom to reach the boundary, while for small $\alpha$ an
atom typically returns many times to the boundary before escaping from
the island. Thus, in the latter limit, the characteristic escape rate
(inverse lifetime $\tau_1^{-1}$) is approximately given by the product
of the probability $2\pi R a/\pi R^2$ for the atom to be at the
boundary and the rate $\alpha D/6a^2$ to overcome the step edge
barrier. Combining these results gives
\begin{equation}
\tau_n(R)=\frac{1}{n}\,
\frac{R^2}{D}\,\left(\kappa_1\frac{a}{\alpha R}+\kappa_2 \right)\,,
\label{taun-eq}
\end{equation}
where $\kappa_1$ and $\kappa_2$ are constants. Indeed, an exact
solution of the corresponding diffusion problem \cite{Harris:1995}
allows one to derive $\tau_n(R)$ exactly in the continuum limit, as we
have shown in Appendix~\ref{taun-app}. In particular, when the escape
is approximated by a Poisson process, one finds $\kappa_1\!\cong\!1$
and $\kappa_2\!=\!1/2$ after proper renormalization and taking into
account the lattice corrections (see Appendix~\ref{taun-app}).  Direct
determination of $\tau_1(R)$ in our simulations confirms this result,
see Fig.~\ref{taun-fig}.

Knowing $\tau_n(R)$ we can calculate the probability
$p_n(R)\!=\!p_n(R(t))$ to find exactly $n$ atoms on top of the island
at time $t$ {\it before} onset of second layer nucleation.  This is
achieved by considering the time evolution of $p_n(R(t))$, which is
described by the master equation
\begin{eqnarray}
\frac{dp_n}{dt}&=&\pi FR(t)^2\,
\bigl[(1\!-\!\delta_{n,0})p_{n-1}-p_n\bigr]\nonumber\\
&&{}+\bigl[\frac{p_{n+1}}{\tau_{n+1}(R(t))}-\frac{p_n}{\tau_n(R(t))}\bigr]
\label{pnmaster-eq}
\end{eqnarray}
with the initial condition $p_n(0)\!=\!\delta_{n,0}$. Note that we
have formally introduced $p_{-1}$ and that $1/\tau_n\!\propto\!n$ so
that the last term on the right hand side of (\ref{pnmaster-eq}) does
not contribute for $n\!=\!0$. As can be expected and is explicitly
shown in Appendix~\ref{taun-app}, the solution of eq.~(\ref{pn-eq}) is
the Poisson distribution
\begin{equation}
p_n(R)=\frac{\bar n(R)^n}{n!}\exp[-\bar n(R)]\,,
\label{pn-eq}
\end{equation}
where the mean number $\bar n(R)$ of atoms on top of the\linebreak
%*****************************************************************
\vspace*{-0.2cm}\noindent
%****************************************************************
island {\it before} onset of nucleation is
\begin{equation}
\bar n(R)=\frac{2\pi}{A^2\Gamma^{\scriptstyle\frac{i}{i\!+\!2}}}
(1\!+\!\tilde\alpha R)^{-\varphi}
       \int_0^R dx\,x^3\,(1\!+\!\tilde\alpha x)^\varphi\,.
\label{barn-eq}
\end{equation}
Here $\tilde\alpha\!\equiv\!\kappa_2\alpha/\kappa_1$ and
$\varphi\!\equiv\!2A^{-2}\Gamma^{2/(i\!+\!2)}/\kappa_2$. An explicit
solution after evaluating the integral in eq.~(\ref{barn-eq}) is given
in eq.~(\ref{barn-app-eq}) of Appendix~\ref{taun-app}. For fixed
$\alpha$ and $\Gamma$, three distinct $R$ regimes can be identified
from eq.~(\ref{barn-eq}): For $\varphi\tilde\alpha R\ll1$ we can use
$(1\!+\!\tilde\alpha x)^\varphi\!\simeq\!1$ in (\ref{barn-eq}), while
for $\alpha R\ll1$ but $\varphi\tilde\alpha R\gg1$ we can use
$(1\!+\!\tilde\alpha x)^\varphi\!\simeq\!\exp(\varphi\tilde\alpha x)$.
For $\tilde\alpha R\!\gg\!1$, we can set $(1\!+\!\tilde\alpha
R)\!\simeq\!\tilde\alpha R$ in (\ref{barn-eq}), and, since the
integral over $x$ is dominated by the upper bound,
$(1\!+\!\tilde\alpha x)\!\simeq\!\tilde\alpha x$ also. We thus obtain
\begin{equation}
\hspace*{-1cm}\bar n(R)\sim\left\{\begin{array}{l@{\hspace*{0.2cm}}l}
\Gamma^{-i/(i\!+\!2)}R^4\,, & 
          R/a\!\ll\!\Gamma^{-2/(i\!+\!2)}\alpha^{-1}\\[0.2cm]
\Gamma^{-1}\alpha^{-1}R^3\,, &
        \Gamma^{-2/(i\!+\!2)}\alpha^{-1}\!\ll\!R/a\!\ll\!\alpha^{-1}
\hspace*{-2cm}\\[0.2cm]
\Gamma^{-1}R^4\,, & \alpha^{-1}\!\ll\!R/a\end{array}\right.
\label{barn-scal-eq}
\end{equation}
The two regimes for large $R$ correspond to a quasi-stationary situation
($dp_n/dt\!=\!0$ in eq.~(\ref{pnmaster-eq})), where $p_n(R)$
from eq.~(\ref{pn-eq}) equals the stationary distribution for $R\!=\!R(t)$
with $\bar n(R)\!=\!\pi FR^2\tau_1(R)$. In these regimes the same result
(\ref{barn-scal-eq}) can be obtained also by integrating
$\rho_1^{\scriptscriptstyle\rm st}$ from eq.~(\ref{rho-st-eq}) over the island
area. In fact, we used this connection to renormalize the constants $\kappa_1$
and $\kappa_2$ in eq.~(\ref{taun-eq}), see Appendix~\ref{taun-app}. The small
$R$ regime in eq.~(\ref{barn-scal-eq}) corresponds to a non-stationary
situation, where $p_n$ in general depends on the function $R(t')$ at all times
$0\!\le\!t'\le t$ and not only on its value $R(t)$ at time $t'\!=\!t$.  This
fact, however, which also concerns the crossover value
$R_\times\!\sim\!\Gamma^{-2/(i\!+\!2)}\alpha^{-1}$ to the non-stationary
regime, is of minor importance here, since we consider the generic growth law
(\ref{R(t)-eq}) throughout the paper. We thus can use $R$ and $t$
interchangeably. Note that the crossover from the non-stationary to the
quasi-stationary situation occurs when $\tau_1(R_\times)\sim \Delta
t(R_\times)$, that means in the non-stationary small $R$ regime the changes in
the radius occur on a faster scale than the escape of an atom from the island,
$\Delta t(R)\ll\tau_1(R)$, while in the two quasi-stationary large $R$ regimes
$\Delta t(R)\gg\tau_1(R)$.

Let us now return to the different scenarios discussed in the
introductory part of this Section. When $\bar n(R)\gtrsim i\!+\!1$,
nucleation of a stable cluster can take place at any instant of time.
The number of nucleations in $\Delta t(R)$ that result from states with
exactly $n$ atoms on top of the island is proportional to
$\omega_n(R)\Delta t(R)$. The total number $n_{\scriptscriptstyle\rm
  nuc}(R)$ is the weighted sum of $\omega_n(R)\Delta t(R)$ over $n$,
i.e.\ we find $n_{\scriptscriptstyle\rm
  nuc}(R)\!=\!\sum_{n=i\!+\!1}^\infty p_n(R)\omega_n(R)\Delta t(R)$
(we are allowed to extend the sum up to infinity due to the sharp
decrease of the Poisson distribution for $n\!\gg\!\bar n(R)$). With
eq.~(\ref{om-nnuc-eq}) we thus obtain for the mean-field nucleation
rate
\begin{eqnarray}
\Omega_{\scriptscriptstyle\rm mf}(R)&=&\sum_{n=i\!+\!1}^\infty
p_n(R)\,\omega_n(R)\nonumber\\
&=&\kappa_{\rm e}\,\frac{D}{a^2}
\left(\frac{\bar n(R)}{\pi R^2}a^2\right)^{i\!+\!1}
\left(\frac{\pi R^2}{a^2}\right)\,.
\label{om-mf-eq}
\end{eqnarray}
Equation~(\ref{om-mf-eq}) can be interpreted as resulting from a local
nucleation rate $\propto D\rho_1^{i\!+\!1}\!=\!D[\bar n(R)/\pi
R^2]^{i\!+\!1}$ integrated over the island area (factor $\pi R^2$).
Compared to the TDT approach the radial variation of the diffusion
profile $\rho_1\!=\!\rho_1(r)$ is neglected in the stochastic
description, so that $\Omega(R)$ from eq.~(\ref{om-tdt-eq}) may be
preferred over eq.~(\ref{om-mf-eq}).\cite{om-mf-comm} However, as will
be discussed further in Sec.~\ref{large-i-subsec} below, for
large $\alpha$ one should use the non-stationary solution of
eqs.~(\ref{diff-eq},\ref{bound-eq}) for calculating $\Omega(R)$ from
(\ref{om-tdt-eq}) corresponding to the small $R$ regime of $\bar n(R)$ in
eq.~(\ref{barn-scal-eq}).

More important, eq.~(\ref{om-mf-eq}) (or (\ref{om-tdt-eq})) can be
used only if $\bar n(R_c)\gtrsim i\!+\!1$ in the relevant time
interval $\Delta t(R_c)$ at the onset of second layer nucleation. The
stochastic description allows us to treat also the fluctuation
dominated case, where $\bar n(R_c)\ll i\!+\!1$. In this situation
$i\!+\!1$ adatoms have to be deposited and to encounter each other on
the island. We can restrict our consideration to the deposition of
exactly $i\!+\!1$ atoms, since for $\bar n(R_c)\ll i\!+\!1$,
fluctuations corresponding to more than $i\!+\!1$ atoms on the island
occur with a probability $\sum_{n=i\!+\!2}^\infty
p_n(R)\!<\!\exp(1)p_{i+1}(R)\bar n(R)/(i\!+\!2)\ll p_{i+1}(R)$.  If an
atom is deposited on the island already containing $i$ atoms, we view
this as the start of a nucleation trial. The number
$n_{\scriptscriptstyle\rm tr}(R)$ of nucleation trials in time $\Delta
t(R)$ is $n_{\scriptscriptstyle\rm tr}(R)\!=\!\pi FR^2\Delta t(R)
p_i(R)$. For a trial to be successful, the $i\!+\!1$ atoms on the
island right after its start have to encounter each other before any
of the atoms escapes by passing the step-edge barrier. The probability
$p_{\scriptscriptstyle\rm enc}(R)$ for
this to happen is
\begin{equation}
p_{\scriptscriptstyle\rm enc}(R)=
1\!-\!\exp[-\omega_{i\!+\!1}(R)\tau_{i\!+\!1}(R)]\,.
\label{pnuc-eq}
\end{equation}
Accordingly, the total number $n_{\scriptscriptstyle\rm nuc}(R)$ of
nucleation events in time $\Delta t(R)$ is now
$n_{\scriptscriptstyle\rm nuc}(R)\!=\!n_{\scriptscriptstyle\rm tr}(R)
p_{\scriptscriptstyle\rm enc}(R)$, and
using eq.~(\ref{om-nnuc-eq}) we obtain for the fluctuation-dominated
nucleation rate
\begin{eqnarray}
\Omega_{\scriptscriptstyle\rm fl}(R)&=&\pi FR^2\, p_i(R)\,
p_{\scriptscriptstyle\rm enc}(R)
\label{om-fl-eq}\\
&=&\pi FR^2 \frac{\bar n(R)^i}{i!}e^{-\bar n(R)}
\Bigl(1\!-\!\exp[-\omega_{i\!+\!1}(R)\tau_{i\!+\!1}(R)]\Bigr)\,.\nonumber
\end{eqnarray}
We note that in both formulae (\ref{om-mf-eq},\ref{om-fl-eq}) the only
parameter not known a priori is the coefficient $\kappa_{\rm e}$,
which has to be taken from simple simulations of the encounter process
(see Fig.~\ref{omn-fig} and the discussion above). Hence they do not
require more input parameters than the expression (\ref{om-tdt-eq})
resulting from the TDT approach.

It remains to clarify, when the mean-field or the fluctuation
dominated situation occurs, i.e.\ when $\Omega_{\scriptscriptstyle\rm
  mf}(R)$ or $\Omega_{\scriptscriptstyle\rm fl}(R)$ has to be used as
second layer nucleation rate. The answer to this question can be found
by self-consistency requirements: Suppose first that the fluctuation
dominated case takes place. Then, using (\ref{om-fl-eq}), one can
calculate the critical radius $R_c$ and check if the condition $\bar
n(R_c)\!\ll\!i\!+\!1$ is fulfilled.  In addition the condition
$\omega_{i+1}(R_c)\Delta t(R_c)\!\gtrsim\!1$ should be fulfilled too,
since the encounter of $i\!+\!1$ atoms in the characteristic time
$\omega_{i+1}(R_c)^{-1}$ should happen before $R_c$ changes. If these
necessary conditions for the fluctuation-dominated case are obeyed,
then the mean-field situation is ruled out. This conclusion can be
drawn, since $\bar n(R)$ is monotonously increasing with $R$, which
implies that $R_c$ following from $\Omega_{\scriptscriptstyle\rm
  fl}(R)$ is always smaller than $R_c$ resulting from
$\Omega_{\scriptscriptstyle\rm mf}(R)$. Hence, when the fluctuations
are likely enough to initiate second layer nucleation, they lead to the
formation of stable clusters at an earlier time $t_c$ than that
expected from the mean-field approach.

We will now show that the fluctuation-dominated case occurs for
$i\!=\!1,2$. The detailed analysis is a bit technical and the reader,
who is interested in the main findings only, may skip the discussion
of the various regimes I-IV in the following subsection and proceed
with the summary of the results given right after this discussion.

\subsection{Small critical nuclei ($i\!=\!1,2$)}
\label{small-i-subsec}
Using $\Omega_{\scriptscriptstyle\rm fl}(R)$ from eq.~(\ref{om-fl-eq})
we can determine the critical radius $R_c$ (or, more precisely,
$R_c'$) by calculating $f_0(t)$ as in the TDT approach (see
eq.~(\ref{f01-eq})). However, for discussing the scaling of $R_c$ with
$\Gamma$ and $\alpha$, it is easier to obtain $R_c$ from the
condition
\begin{equation}
\Omega_{\scriptscriptstyle\rm fl}(R_c)\Delta t(R_c)\simeq 1\,,
\label{rcdet-eq}
\end{equation}
which expresses the fact that the probability of second layer
nucleation in $\Delta t(R_c)$ becomes of the order of one. Since we
consider the fluctuation-dominated case for small critical nuclei here
($i\!=\!1,2$), we assume $\bar n(R_c)\!\ll\!1$ and thus set
$\exp[-\bar n(R_c)]\!\simeq\!1$, when inserting
$\Omega_{\scriptscriptstyle\rm fl}(R_c)$ from eq.~(\ref{om-fl-eq})
into eq.~(\ref{rcdet-eq}).

Four different regimes are then predicted by eq.~(\ref{rcdet-eq}):
\begin{list}{}{\setlength{\topsep}{0.1cm}\setlength{\leftmargin}{0cm}
    \setlength{\labelwidth}{-0.2cm}\setlength{\labelsep}{0.2cm}
    \setlength{\itemsep}{0.1cm}\setlength{\parsep}{0cm}}
\item[\sl Regime I:] In the limit $\alpha\!\to\!0$ we have $\bar
  n(R)\!\sim\!\Gamma^{-i/(i\!+\!2)}R^4$ and
  $\tau_{i\!+\!1}\!\to\!\infty$.  Hence we obtain from
  eqs.~(\ref{deltat-eq},\ref{om-fl-eq},\ref{rcdet-eq})
  $FR_c^2\Gamma^{-i^2/(i\!+\!2)}R_c^{4i}F^{-1}\Gamma^{-i/(i\!+\!2)}R_c^2
  \!\sim\!R_c^{4(i\!+\!1)}\Gamma^{-i(i\!+\!1)/(i\!+\!2)}\!\sim\!{\rm
    const.}$, i.e.\ 
  \begin{equation}
  R_c\sim\Gamma^{i/[4(i\!+\!2)]}
  \label{rc-I-eq}
  \end{equation}
  From (\ref{rc-I-eq}) follows $\bar n(R_c)\sim {\rm const.}$, which
  means that the assumption of a fluctuation-dominated situation is
  not necessarily justified. In fact, eq.~(\ref{rc-I-eq}) appears here
  as the result of a rather lengthy calculation, but in the limit
  $\alpha\!\to\!0$, the same scaling behavior (\ref{rc-I-eq}) can be
  obtained very simply by calculating the average time needed for the
  deposition of $i\!+\!1$ atoms (see
  ref.~\onlinecite{Rottler/Maass:1999}). Hence, despite $\bar
  n(R_c)\!\simeq\!i\!+\!1$, eq.~(\ref{rc-I-eq}) gives the correct
  scaling behavior. However, eq.~(\ref{rc-I-eq}) predicts
  $\omega_{i+1}(R_c)\Delta
  t(R_c)\!\sim\!DR_c^{-2i}F^{-1}\Gamma^{-i/(i\!+\!2)}R_c^2\!\sim\!
  \Gamma^{-(i^2\!-\!i\!-\!4)/[2(i\!+\!2)]}$ and since
  $\Gamma\!=\!D/Fa^4\!\gg\!1$,\cite{gamma-comm} the inequality
  $\omega_{i+1}(R)\Delta t(R_c)\!\gtrsim\!1$ becomes violated for
  $i\!\ge\!3$.  For $i\!\ge\!3$ therefore, the condition
  $\omega_{i+1}(R_c)\Delta t(R_c)\!\sim\!1$ should be used for
  calculating $R_c$, and because this yields $\bar
  n(R_c)\!>\!i\!+\!1$, one may alternatively use
  $\Omega_{\scriptscriptstyle\rm mf}(R_c)\Delta t(R_c)\simeq1$ as the
  determining relation (see Sec.~\ref{large-i-subsec}).
  
\item[\sl Regime II:] With increasing $\alpha$, for $i\!\le\!2$,
  either the non-stationarity condition $\tau_1(R_c)\!\gg\!\Delta
  t(R_c)$ ($\bar n(R_c)\!\sim\!\Gamma^{-i/(i\!+\!2)}R_c^4$ in
  eq.~(\ref{om-fl-eq})) or the condition
  $\omega_{i+1}(R_c)\tau_{i+1}(R_c)\!\gg\!1$ ($p_{\scriptscriptstyle
    enc}\simeq 1$ in eq.~(\ref{om-fl-eq})) breaks down first. Taking
  $R_c$ from eq.~(\ref{rc-I-eq}), the first condition implies
  $\alpha\!\ll\!\Gamma^{-(i\!+\!8)/[4(i\!+\!2)]}$, while the second
  implies $\alpha\!\ll\!\Gamma^{-i(2i\!-\!1)/[4(i\!+\!2)]}$.  Since
  the first condition is more restrictive for $i\!\le\!2$, regime I
  ceases to be valid when $\alpha$ becomes larger than
  $\Gamma^{-(i\!+\!8)/[4(i\!+\!2)]}$ and the quasi-stationarity
  situation is reached. In eq.~(\ref{om-fl-eq}) we now have to take
  $\bar n(R_c)\!=\!\pi FR^2\tau_1(R)\!\sim\!\Gamma^{-1}\alpha^{-1}R^3$
  (see eq.~(\ref{barn-scal-eq})) and it follows
  $\Omega_{\scriptscriptstyle\rm fl}(R_c)\Delta t(R_c)\!\sim\!
  \alpha^{-i}\Gamma^{-i(i\!+\!3)/(i\!+\!2)}
  R_c^{3i\!+\!4}\!\sim\!{\rm const.}$, i.e.\ 
  \begin{equation}
  R_c\sim\alpha^{i/(3i\!+\!4)}\Gamma^{i(i\!+\!3)/[(i\!+\!2)(3i+4)]}\,.
  \label{rc-II-eq}
  \end{equation}
  Since $\bar
  n(R_c)\!\sim\!(\Gamma^{-(i\!+\!8)/4(i\!+\!2)}\alpha^{-1})^{4/(3i+4)}\!\ll\!1$
  the condition for a fluctuation-dominated situation is fulfilled,
  and since $\Delta t(R_c)\!\gg\!\tau_1(R_c)\!\simeq\!\tau_{i+1}(R_c)$
  and $\omega_{i+1}(R_c)\tau_{i+1}(R_c)\!\gg\!1$ the condition
  $\omega_{i+1}(R_c)\Delta t(R_c)\!\gtrsim\!1$ is obeyed too.
     
\item[\sl Regime III:] By further increasing $\alpha$ we obtain
  $\omega_{i+1}(R_c)\tau_{i+1}(R_c)\!\ll\!1$ for
  $\alpha\!\gg\!\Gamma^{-i(i\!+\!3)(2i\!-\!1)/[2(i\!+\!2)(i^2\!+\!i\!+\!2)]}$.
  Hence we now have to use $p_{\scriptscriptstyle
    enc}\!\simeq\!\omega_{i+1}(R_c)\tau_{i+1}(R_c)$ when inserting
  eq.~(\ref{om-fl-eq}) into eq.~(\ref{rcdet-eq}) and find
  \begin{equation}
  R_c\sim\alpha^{(i\!+\!1)/(i\!+\!5)}
              \Gamma^{i(i\!+\!3)/[(i\!+\!2)(i+5)]}\,.
  \label{rc-III-eq}
  \end{equation}
  The condition $\bar
  n(R_c)\!\sim\!\Gamma^{-1}\alpha^{-1}R_c^3\!\ll\!1$ requires
  $\alpha^{i\!-\!1}\!\ll\!\Gamma^{-(i^2\!+\!i\!-\!5)/(i\!+\!2)}$ and
  is fulfilled for $i\!=\!1$. For $i\!=\!2$, it is valid for
  $\alpha\!\ll\!\Gamma^{-1/4}\!\sim\!a/R_c$. The second requirement
  $\omega_{i+1}(R_c)\Delta t(R_c)\!\gtrsim\!1$ gives
  $\alpha^{i\!-\!1}\!\ll\!\Gamma^{-(i^3\!+\!2i^2\!-\!4i\!-\!5)/
    [(i\!+\!1)(i\!+\!2)]}$ and again is obeyed for $i\!=\!1$ and valid
  for $i\!=\!2$ as long as $\alpha\!\ll\!\Gamma^{-1/4}\!\sim\!a/R_c$.
  
\item[\sl Regime IV:] In this last regime $\alpha$ becomes larger than
  $a/R_c$, that means eq.~(\ref{rc-III-eq}) predicts the regime to
  occur for $\alpha\!\gg\!\Gamma^{-i/[2(i\!+\!2)]}$.  Taking $\bar
  n(R_c)\!\sim\!\Gamma^{-1}R_c^4$ from eq.~(\ref{barn-scal-eq}) and
  $\omega_{i+1}(R_c)\tau_{i+1}(R_c)\!\sim\!R_c^{-2(i-1)}$ from
  eqs.~(\ref{omn-eq},\ref{taun-eq}), we find
  \begin{equation}
  R_c\sim\Gamma^{i/[2(i\!+\!2)]}\,.
  \label{rc-IV-eq}
  \end{equation}
  We used $\omega_{i+1}(R_c)\tau_{i+1}(R_c)\!\ll\!1$
  ($p_{\scriptscriptstyle\rm enc}(R_c)\!\ll\!1$) to derive (\ref{rc-IV-eq}),
  which for $i\!=\!2$ is valid and for $i\!=\!1$ is obeyed when taking into
  account the prefactors (for $i\!=\!1$
  $\omega_{i+1}(R_c)\tau_{i+1}(R_c)\!=\!\kappa_{\rm e}\kappa_2$). Moreover,
  eq.~(\ref{rc-IV-eq}) gives $\bar n(R_c)\sim\Gamma^{(i\!-\!2)/(i\!+\!2)}$,
  which is much smaller than one for $i\!=\!1$. For $i\!=\!2$, a decision on
  whether the fluctuation-dominated or the mean-field situation occurs would
  require a closer inspection of the prefactors. However, since for
  $\alpha\!\gg\!a/R_c$ one finds the same scaling (\ref{rc-IV-eq}) in the
  mean-field situation (see Sec.~\ref{large-i-subsec}), eq.~(\ref{rc-IV-eq})
  is valid in any case. The second condition $\omega_{i+1}(R_c)\Delta
  t(R_c)\!\gtrsim\!1$ is fulfilled for $i\!=\!1$, and for $i\!=\!2$ the
  situation again depends on the prefactors.
\end{list}

In summary we have found that the second layer nucleation for $i\!=\!1,2$
occurs due to various mechanisms in four distinct regimes I-IV: In regime I
($\alpha\!\ll\!\Gamma^{-(i\!+\!8)/[4(i\!+\!2)]}$), the nucleation takes place
once $i\!+\!1$ have been deposited on the island, in regime II
($\Gamma^{-(i\!+\!8)/[4(i\!+\!2)]}\!\ll\!\alpha\!\ll\!
\Gamma^{-i(i\!+\!3)(2i\!-\!1)/[2(i\!+\!2)(i^2\!+\!i\!+\!2)]}$) the loss of
atoms becomes important and the nucleation takes place once the probability
for finding $i\!+\!1$ atoms on the island at some time instant in $\Delta
t(R)$ becomes of the order of one, in regime III
($\Gamma^{-i(i\!+\!3)(2i\!-\!1)/[2(i\!+\!2)(i^2\!+\!i\!+\!2)]}
\!\ll\!\alpha\!\ll\!\Gamma^{-i/[2(i\!+\!2)]}$) the probability
$p_{\scriptscriptstyle\rm enc}$ for the encounter of $i\!+\!1$ atoms during a
nucleation trial has to be taken into account in addition to the probability
for the occurrence of $i\!+\!1$ atoms, and in regime IV
($\alpha\!\gg\!\Gamma^{-i/[2(i\!+\!2)]}$) both the occurrence and encounter
probability matter but these probabilities no longer depend on the step edge
barrier. For convenient reference, we provide the exponents $\gamma$ and $\mu$
defined in eq.~(\ref{gamma-mu-eq}) and their corresponding ranges of validity
in Table~\ref{small-i-tab}.  When comparing the scaling in the
fluctuation-dominated situation with that predicted by
eqs.~(\ref{gamma-tdt-eq},\ref{mu-tdt-eq}) of the TDT approach it is remarkable
that the same behavior is found in regime III and IV. We believe this to be
caused by the fortunate circumstance that local nucleation rates of form
$\propto D\rho_1^{i\!+\!1}$ might be {\it effectively} applicable even if the
requirements for the mean-field situation are not fulfilled (see
however \onlinecite{Kallabis/etal:1998}).

Moreover, we have to note that for $i\!=\!1$ the lower and upper
crossovers $\alpha_2$ and $\alpha_3$ specifying regime III (see
table~\ref{small-i-tab}) both scale as $\Gamma^{-1/6}$. This is
due to the fact that when $\omega_{i+1}(R_c)\tau_{i+1}(R_c)$ becomes
less than one, we already obtain $\alpha\!\gg\!a/R_c$, which is the
condition for regime IV.  Nevertheless, due to the pronounced small
$R$ corrections to eq.~(\ref{omn-eq}) for $i\!=\!1$ (see
Fig.~\ref{omn-fig}) both conditions
$\omega_{i+1}(R_c)\tau_{i+1}(R_c)\!\ll\!1$ and $\alpha\!\ll\!a/R_c$
can be fulfilled in a small transient regime III. However, for
$i\!=\!1$ this is no longer a true scaling regime, where a simple
power law dependence of $R_c$ on $\Gamma$ and $\alpha$ can be
identified. In this respect, the small $\alpha$ regime of the TDT
approach does not occur for $i\!=\!1$, also not at larger $\alpha$.

\subsection{Comparison with simulations for i=1,2}
\label{comp-fl-sec}
\begin{figure}[b]
\begin{center}\epsfig{file=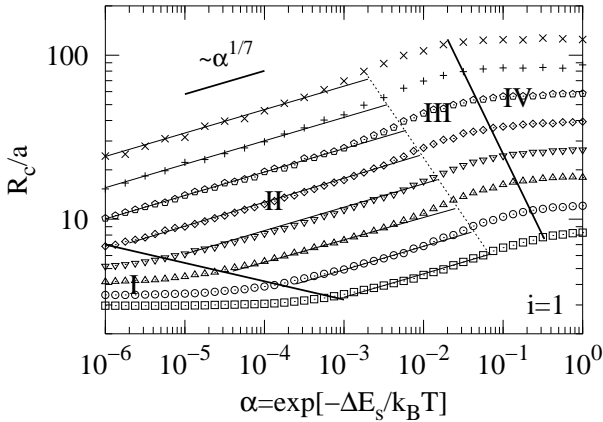,width=8.5cm}\end{center}
\vspace*{-0.1cm}
\caption{Critical island size $R_c$ as a function of $\alpha$ obtained from
  simulation of the single-island model for 8 different $\Gamma$
  values starting from $10^5$ ($\square$) and ending at $10^{12}$
  ($\times$). Between these values, $\Gamma$ is increased by a factor
  of 10. The various regimes I-IV are indicated together with the
  border line with slope (-1/9) between regimes I and II and the
  border line with slope (-1) between regimes III and IV. The dashed
  border line separating the transient regime III from regime II was
  calculated numerically from the condition
  $\omega_2(R_c)\tau_2(R_c)\simeq1$ (see text).}
\label{rc-i1-fig}
\end{figure}

\vspace*{0.1cm}
Taking $\Omega_{\rm\scriptscriptstyle fl}$ from eq.~(\ref{om-fl-eq})
we can calculate $f_0(t)$ according to eq.~(\ref{f01-eq}).
Representative results for $i\!=\!1$ are shown in Fig.~\ref{f-t-fig}
(solid lines), and the comparison with the Monte Carlo data yields a
very good agreement. The $R_c$ values 
%****************************************************************
\begin{table}[b]
\caption{The exponents $\gamma$ and $\mu$ characterizing the scaling 
$R_c\!\sim\!\Gamma^\gamma\alpha^\mu$ in the various regimes I-IV for 
$i\!=\!1,2$}\vspace*{0.1cm}
\begin{tabular}{cccc}
Regime & Range\tablenote{The crossover values scale as
$\alpha_k\!\sim\!\Gamma^{-\delta_k}$ where
$\delta_1\!\equiv\!(i\!+\!8)/4(i\!+\!2)$,
$\delta_2\!\equiv\!i(i\!+\!3)(2i\!-\!1)/2(i\!+\!2)(i^2\!+\!i\!+\!2)$,
and $\delta_3\!\equiv\!i/2(i\!+\!2)$.} & $\gamma$ & $\mu$ \\\tableline
I & $0\!\le\!\alpha\!\ll\!\alpha_1$ & $i/4(i\!+\!2)$ & 0\\
II & $\alpha_1\!\ll\!\alpha\!\ll\!\alpha_2$ 
                 & $i(i\!+\!3)/(i\!+\!2)(3i\!+\!4)$ & $i/(3i\!+\!4)$\\
III & $\alpha_2\!\ll\!\alpha\!\ll\!\alpha_3$ 
               & $i(i\!+\!3)/(i\!+\!2)(i\!+\!5)$ & $(i\!+\!1)/(i\!+\!5)$\\
IV & $\alpha_3\!\ll\!\alpha\!\le\!1$ & $i/2(i\!+\!2)$ & 0\\
\end{tabular}
\label{small-i-tab}
\end{table}\noindent
%****************************************************************
derived from $f_0(t)$ are
plotted as a function of $\alpha$ for $\Gamma$
val-
ues in the range $10^5-10^8$ in Fig.~\ref{rc-i1-fig}. Note that,
compared to the results of the full island model shown in
Fig.~\ref{rc-I-II-fig}, the data cover the full $\alpha$ range from
zero to one, since the restrictions imposed by island coalescence in
the multi-island model are not present in the single-island model (see
also the discussion in Sec.~\ref{sim-sec} above). Moreover the simpler
single-island model allows one to explore the behavior for larger
$\Gamma$ values in the range $\Gamma=10^9-10^{12}$ also.  It is
possible to fit the $R_c$ curves over the entire range of $\alpha$ and
$\Gamma$ values (see Sec.~\ref{negligible-subsec}) but we focus on the
scaling behavior of $R_c$ in the following in order to demonstrate the
various scaling regimes associated with the different physical
mechanisms of second layer nucleation.

\vspace*{0.2cm}
Indeed, the simulated data in Fig.~\ref{rc-i1-fig} confirm the theoretical
predictions. For small $\alpha\!\ll\!\alpha_1$ (regime I), $R_c$ is
independent of $\alpha$, while for
$\alpha\!\gg\!\alpha_1\!\sim\!\Gamma^{-3/4}$ (regime II) we find
$R_c\sim\Gamma^{4/21}\alpha^{1/7}$. Since $R_c(\alpha_1)\sim\alpha_1^{-1/9}$
at the crossover, the boundary line between regimes I and II
has slope (-1/9).  The correctness of the scaling of $R_c$ with $\Gamma$ in
regimes I and II can be deduced from the offset of the curves for various
$\Gamma$ in Fig.~\ref{rc-i1-fig} (alternatively, one can collapse the data
onto a common master curve by a proper rescaling as it was shown in
ref.~\onlinecite{Rottler/Maass:1999}).  Regime II is followed by the transient
regime III, and the dashed border line separating regime III from regime II
was determined numerically from the condition
$\omega_2(R_c)\tau_2(R_c)\simeq1$ by using the results for $\omega_2(R_c)$ and
$\tau_2(R_c)$ displayed in Figs.~\ref{omn-fig},\ref{taun-fig}. For
$\alpha\gg\Gamma^{-i/[2(i\!+\!2)]}$ (regime IV), we find
$R_c\!\sim\!\Gamma^{i/[2(i\!+\!2)]}$ independent of $\alpha$, and the boundary
line between regimes III and IV has slope (-1).

%****************************************************************
\noindent
%**************************************************************************
\begin{figure}[t]
\begin{center}\hspace*{-0.2cm}
\epsfig{file=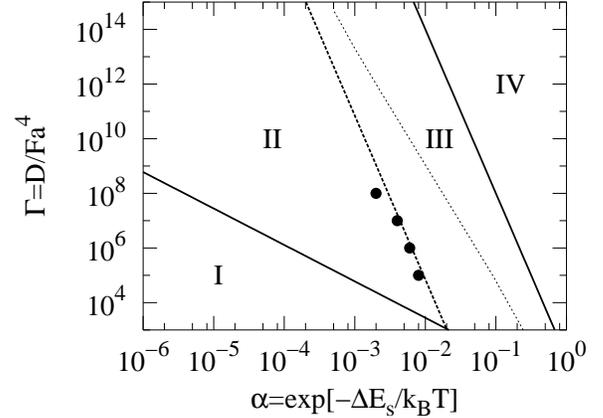,width=8.5cm}
\end{center}
\vspace*{-0.2cm}
\caption{The various regions characterizing the mechanism of second
  layer nucleation for $i\!=\!1$ in an $\alpha-\Gamma$ diagram. The
  thick dashed line with slope (-6) marks the onset of layer-by-layer
  growth (the circles refer to the onset of island coalescence
  obtained in the full simulation, see Fig.~\ref{rc-I-II-fig}).}
\label{diagram-fig}
\end{figure}\noindent
%***************************************************************************
Figure~\ref{diagram-fig} depicts the various regions characterizing
the mechanism of second layer nucleation for $i=1$ in an
$\alpha-\Gamma$ diagram. Varying $\Gamma$ and $\alpha$ within one of
the regions results in the corresponding behavior of $R_c$ according
to eqs.~(\ref{rc-I-eq},\ref{rc-II-eq},\ref{rc-IV-eq}). The border line
between regions I and II has slope $(-4/3)$, between regions III and
IV slope $(-6)$, and the dashed line marks the border line between
regions II and III.  In addition, we have drawn the transition line
from rough multilayer to smooth layer-by-layer growth into the
diagram. In our simulations island coalescence occurs in regime II
(see Fig.~\ref{rc-I-II-fig}), where
$R_c\!\sim\!\Gamma^{4/21}\alpha^{1/7}$.  The criterion
$R_c\!\simeq\!l\!\sim\!\rho_x^{-1/2}\!\sim\!\Gamma^{1/6}$ thus yields
$\alpha_\star(\Gamma)\!\sim\!\Gamma^{-1/6}$.

\begin{figure}[t]
\begin{center}\epsfig{file=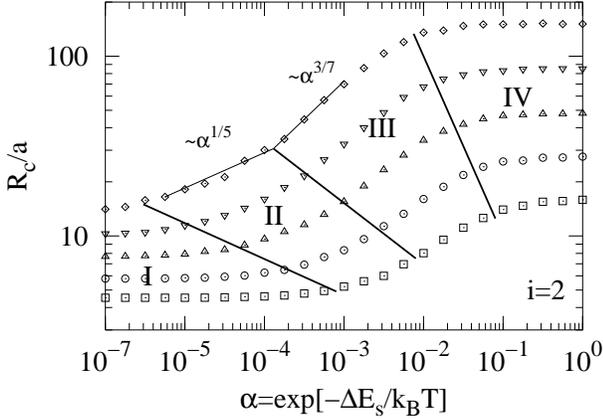,width=8.5cm}\end{center}
\vspace*{-0.2cm}
\caption{Critical island size $R_c$ for $i\!=\!2$ obtained from the
  single-island model for 5 different $\Gamma$ values between $10^5$
  ($\scriptscriptstyle\square$) and $10^9$ ($\diamond$). Between these values,
  $\Gamma$ is increased by a factor of 10 as in Fig.~5. The scaling
  regimes I-IV are indicated together with the respective border lines
  with slope (-1/5) between regimes I and II, slope (-1/3) between
  regime II and III and slope (-1) between regimes III and IV.}
\label{rc-i2-fig}
\end{figure}

Results for $R_c$ obtained from simulations for a critical nucleus of
size $i=2$ are shown in Fig.~\ref{rc-i2-fig}. Again the results
confirm the predictions of the theory. In particular, for large
$\Gamma$, the exponents $\mu=1/5$ in regime II and $\mu=3/7$ in regime
III can be clearly identified. In contrast to the behavior for
$i\!=\!1$ shown in Fig.~\ref{rc-i1-fig}, regime III develops into a
full scaling regime.

\vspace*{-0.3cm}
\subsection{Large critical nuclei ($i\!\ge\!3$)}
\label{large-i-subsec}\vspace*{-0.2cm}
Analogous to the fluctuation-dominated case treated in the previous
subsection we can obtain the scaling of 
$R_c$ with $\Gamma$ and $\alpha$ from the condition
$\Omega_{\scriptscriptstyle\rm mf}(R_c)\Delta t(R_c)\simeq1$ with
$\Omega_{\scriptscriptstyle\rm mf}(R)$ and $\Delta t(R_c)$ from
eqs.~(\ref{om-mf-eq},\ref{deltat-eq}), respectively. For critical
island radii belonging to the two quasi-stationary large $R$-regimes
in eq.~(\ref{barn-scal-eq}) this gives the same behavior
(\ref{gamma-tdt-eq},\ref{mu-tdt-eq}) as predicted by the TDT approach.
However, for large step edge barriers corresponding to the
non-stationary small $\alpha$-regime in eq.~(\ref{barn-scal-eq}) we
find $\Omega_{\scriptscriptstyle\rm mf}(R_c)\Delta t(R_c)\!\sim\!
D(\Gamma^{-i/(i\!+\!2)}R_c^4)^{i\!+\!1}R_c^{-2i}F^{-1}\Gamma^{-i/(i\!+\!2)}
R_c^2\!\sim\!R_c^{2(i\!+\!3)}\Gamma^{-(i\!-\!1)}\!\sim\!{\rm const.}$,
i.e.\ \vspace*{-0.2cm}
\begin{equation}
R_c\sim\Gamma^{(i\!-\!1)/[2(i\!+\!3)]}\,.
\label{rcI-tdt-eq}
\end{equation}
%***************************************************************************
\vspace*{-0.2cm}\noindent
%****************************************************************
With increasing $\alpha$ this scaling breaks down when $R_c$ enters
the quasi-stationary regime in eq.~(\ref{barn-scal-eq}) that means for
$\alpha\!\gg\!\Gamma^{-2/(i\!+\!2)}R_c^{-1}\!\sim\!
\Gamma^{-(i^2\!+\!5i\!+\!10)/[2(i\!+\!2)(i\!+\!3)]}$.

\begin{figure}[t]
\begin{center}\epsfig{file=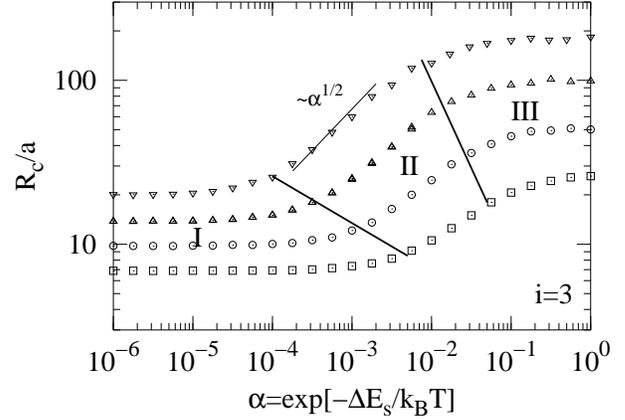,width=8.5cm}\end{center}
\vspace*{-0.2cm}
\caption{Critical island size $R_c$ for $i=3$ obtained from the
  single-island model for 4 different $\Gamma$ values between $10^5$
  ($\scriptscriptstyle\square$) and $10^{8}$ ($\triangledown$). The
  scaling regimes I-III are indicated together with the respective
  border lines with slope (-5/17) between regimes I and II and slope
  (-1) between regimes II and III.}
\label{rc-i3-fig}
\end{figure}

For $i\!\ge\!3$ we thus have in total three distinct regimes I-III
with different mechanisms for second layer nucleation: In regime I
($\alpha\!\ll\!\Gamma^{-(i^2\!+\!5i\!+\!10)/[2(i\!+\!2)(i\!+\!3)]}$)
the nucleation takes place once the island radius $R$ has grown large
enough so that the encounter of $i\!+\!1$ atoms out of typically $\bar
n(R)\gtrsim i\!+\!1$ atoms happens in a time comparable to $\Delta
t(R)$, in regime II $\bar n(R)$ becomes dependent on $\alpha$, while
in regime III, for large $\alpha\!\gg\!a/R_c$, $\tau_1(R)$ no longer
depends on the step edge barrier and $\bar n(R)$ becomes independent
of $\alpha$ again. The overall behavior characterized by the scaling
exponents $\gamma$ and $\mu$ is summarized in Table~\ref{large-i-tab}.
Computer simulations for $i\!=\!3$ are in accordance with these
theoretical predictions, see Fig.~\ref{rc-i3-fig}. The predicted
scaling $R_c\!\sim\!\alpha^{1/2}$ in regime II is not yet fully
developed for the $\Gamma$-values in the range $10^5-10^8$ but it can
be expected to become more clearly visible for larger $\Gamma$.
However, we could not obtain reliable simulation results for larger
$\Gamma$ values, since the amount of CPU time for determining the
onset of second layer nucleation becomes tremendous due to the
increasing number of atoms contributing to the nucleation event.

\subsection{Influence of metastable clusters}
\label{simple-meta-subsec}
To demonstrate how the presence of metastable nuclei may be included
into the general procedure presented in
Sec.~\ref{simple-proc-subsec}, we consider, as in
ref.~\onlinecite{Krug/etal:2000}, the simplest case of second layer
nucleation of a trimer ($i\!=\!2$), when a dimer is metastable with
characteristic dissociation time $\tau_{\rm\scriptscriptstyle dis}$.
For $i\!=\!2$ we have to deal with the fluctuation-dominated
situation. We note in passing that this can be true even
%****************************************************************
\vspace*{-0.2cm}\noindent
%****************************************************************
for larger
$i$ when metastable clusters can form, since their presence tends to
drive second layer nucleation into the fluctuation-dominated
situation.

In contrast to the discussion leading to eq.~(\ref{om-fl-eq}) for the
non-interacting particle model, the formation of the stable trimer is
not necessarily the rate limiting process. It is possible that the
dissociation time $\tau_{\rm\scriptscriptstyle dis}$ becomes so large
that the nucleation happens effectively instantaneously once the dimer
has formed. To decide whether the formation of the stable trimer or
metastable dimer is rate limiting, we have to compare
$p_1(R)p_{\scriptscriptstyle\rm enc}^{\scriptscriptstyle\rm (2)}(R)$
with $p_2(R) p_{\scriptscriptstyle\rm enc}^{\scriptscriptstyle\rm
  (3)}(R)$, where $p_{\scriptscriptstyle\rm
  enc}^{\scriptscriptstyle\rm (j)}(R)$ denotes the encounter
probability of $j$ atoms (in Sec.~\ref{simple-proc-subsec} no
superscript (j) was introduced, since only $j\!=\!i\!+\!1$ had to be
considered). Hence we write
\begin{equation}
\Omega_{\scriptscriptstyle\rm fl}(R)\!=\!
\left\{\begin{array}{l@{\hspace{0.4cm}}l}
\pi FR^2 
   p_2(R) p_{\scriptscriptstyle\rm enc}^{\scriptscriptstyle\rm (3)}(R)\,, &
p_1 p_{\scriptscriptstyle\rm enc}^{\scriptscriptstyle\rm (2)}\!\gg\!
p_2 p_{\scriptscriptstyle\rm enc}^{\scriptscriptstyle\rm (3)} \\[0.3cm]
\pi FR^2
   p_1(R) p_{\scriptscriptstyle\rm enc}^{\scriptscriptstyle\rm (2)}(R)\,, &
p_1 p_{\scriptscriptstyle\rm enc}^{\scriptscriptstyle\rm (2)}\!\ll\!
p_2 p_{\scriptscriptstyle\rm enc}^{\scriptscriptstyle\rm (3)}\end{array}\right.
\label{om-fl-meta-eq}
\end{equation}

\vspace*{-0.1cm} To calculate the occupation probabilities $p_n(R)$,
$n\!\le\!3$, we first need to know the modified lifetimes $\tau_n'(R)$
of states with exactly $n$ atoms on top of the island. Clearly,
$\tau_1'(R)\!=\!\tau_1(R)$ with $\tau_1(R)$ from eq.~(\ref{taun-eq}),
since the metastable dimer has no influence on the lifetime of a
single atom. The characteristic time $\tau_2'(R)$, however, will be
enlarged in comparison to $\tau_2(R)$ from eq.~(\ref{taun-eq}) and can
be estimated as follows (we disregard any prefactors): As in
ref.~\onlinecite{Krug/etal:2000} we consider the first deposited atom
as\vspace*{-0.35cm}\linebreak
%****************************************************************
\vspace*{-0.15cm}
\begin{table}
\caption{The exponents $\gamma$ and $\mu$ characterizing the scaling 
$R_c\!\sim\!\Gamma^\gamma\alpha^\mu$ in the various regimes I-III for 
$i\!\ge\!3$}\vspace*{0.1cm}
\begin{tabular}{cccc}
Regime & Range\tablenote{The crossover values scale as
$\alpha_k\!\sim\!\Gamma^{-\delta_k}$, where
$\delta_1\!\equiv\!(i^2\!+\!5i\!+\!10)/2(i\!+\!2)(i\!+\!3)$,
and $\delta_2\!\equiv\!i/2(i\!+\!2)$.} & $\gamma$ & $\mu$ \\\tableline
I & $0\!\le\!\alpha\!\ll\!\alpha_1$ & $(i\!-\!1)/2(i\!+\!3)$ & 0\\
II & $\alpha_1\!\ll\!\alpha\!\ll\!\alpha_2$ 
               & $i(i\!+\!3)/(i\!+\!2)(i\!+\!5)$ & $(i\!+\!1)/(i\!+\!5)$\\
III & $\alpha_2\!\ll\!\alpha\!\le\!1$ & $i/2(i\!+\!2)$ & 0\\
\end{tabular}
\label{large-i-tab}
\end{table}\noindent
%****************************************************************
immobile and the second deposited atom as diffusing.  Once the
second atom has been deposited it needs a time of order
$(R^2/D+\tau_{\rm\scriptscriptstyle dis})$ to reach the step edge,
since one encounter with the first deposited atom typically takes
place during one traversal of the island within time
$\tau_{\rm\scriptscriptstyle tr}\!\equiv\!R^2/D$.
\cite{Krug/etal:2000} At the boundary the second atom is ``reflected''
a typical number $M\!\sim\!\alpha^{-1}$ of times before leaving the
island.  Between all reflections, the overall elapsed time is of order
$(MRa/D+m\tau_{\rm\scriptscriptstyle dis})$, where $Ra/D$ is the
typical time for a single atom to return to the edge and
$m\!\sim\!(MRa/D)/\tau_{\rm\scriptscriptstyle tr}\!\sim\!a/\alpha R$
is the typical number of times the second atom encounters the first
atom.\cite{Krug/etal:2000} Summing up all time contributions we obtain
(neglecting the prefactors belonging to the four individual terms)
\begin{equation}
\tau_2'(R)
\sim(\tau_{\rm\scriptscriptstyle
  tr}+\tau_{\rm\scriptscriptstyle dis})\!+\!\frac{a}{\alpha
R}(\tau_{\rm\scriptscriptstyle tr}+\tau_{\rm\scriptscriptstyle dis})\,.
\label{tau2-prime-eq}
\end{equation}
Note that for $\tau_{\rm\scriptscriptstyle
  dis}\!\ll\!\tau_{\rm\scriptscriptstyle tr}\!=\!R^2/D$, $\tau_2'(R)$
reduces to $\tau_2(R)$ from eq.~(\ref{taun-eq}) (without prefactors).
To estimate $\tau_3'(R)$ we note that if the dimer state is the
prevalent one, $\tau_3'(R)\!\sim\!\tau_1(R)$, whereas, if all three
atoms are likely to be separated, $\tau_3'(R)\!\sim\!\tau_3(R)$. Since
$\tau_3(R)\!\sim\!\tau_1(R)$, we find $\tau_3'(R)\!\sim\!\tau_1(R)$ in
either case. In the strong barrier limit $\alpha\!\ll\!a/R$, in
particular, the first two terms on the right hand side of
eq.~(\ref{tau2-prime-eq}) can be neglected, and, since
$\tau_1\sim\!Ra/D\alpha\!\sim\!\tau_{\rm\scriptscriptstyle tr}a/\alpha
R$, we can simply write
$\tau_2'\sim\tau_1(1+\tau_{\rm\scriptscriptstyle
  dis}/\tau_{\rm\scriptscriptstyle tr})$ which agrees with the result
derived in ref.~\onlinecite{Krug/etal:2000}. This finding for strong step
edge barriers implies that for $\tau_{\rm\scriptscriptstyle
  dis}/\tau_{\rm\scriptscriptstyle tr}\!\gg\!1$ the two atoms are
effectively always in the dimer state and
$\tau_2'\!\sim\!\tau_2\tau_{\rm\scriptscriptstyle
  dis}/\tau_{\rm\scriptscriptstyle tr}$, while for
$\tau_{\rm\scriptscriptstyle dis}/\tau_{\rm\scriptscriptstyle
  tr}\!\ll\!1$ they are effectively always separated and
$\tau_2'\!\sim\!\tau_2$.

When inserting the modified lifetimes into eq.~(\ref{pnmaster-eq}) and
neglecting states with $n\!>\!i\!+\!1\!=\!3$ ($p_n\!\simeq\!0$ for
$n\!>\!3$ until onset of nucleation in the fluctuation-dominated
situation), we can calculate the occupation probabilities $p_n(t)$.
In the quasi-stationary limit ($dp_n/dt\!=\!0$ but $R=R(t)$),
in particular, we obtain (for $0\!\le\!n\!\le\!i\!+\!1\!=\!3$)
\begin{equation}
p_n=\frac{\prod_{j=1}^n q_j}{\sum_{j=0}^{i+1}\prod_{k=1}^j q_k}\,,
\hspace*{0.3cm}q_j\!\equiv\!\pi FR^2\tau_j'\,.
\label{pn-meta-eq}
\end{equation}

To calculate the encounter probabilities $p_{\scriptscriptstyle\rm
  enc}^{\scriptscriptstyle\rm
  (n)}(R)\!=\!1-\exp[-\omega_n'(R)\tau_n'(R)]$, $n\!=\!2,3$, we
furthermore need to know the modified encounter rates $\omega_n'(R)$.
{}From eq.~(\ref{weff-eq}) in Appendix~\ref{omn-app} we find
$\omega_2'\!=\!w_1$ (eq.~(\ref{weff-eq}) for $i\!=\!1$) and
$\omega_3'\!\simeq\!w_1w_2/v_2'$ (eq.~(\ref{weff-eq}) for $i\!=\!2$),
where $w_1\!\sim\!w_2\!\sim\!\tau_{\scriptscriptstyle\rm tr}^{-1}$
from eq.~(\ref{ws-all-eq}) and $v_2'\!=\!\tau_{\scriptscriptstyle\rm
  dis}^{-1}$ (modification of eq.~\ref{vs-eq}), i.e.\ 
$\omega_2'\!\sim\!1/\tau_{\scriptscriptstyle\rm tr}$ and
$\omega_3'\!\sim\!\tau_{\scriptscriptstyle\rm
  dis}/\tau_{\scriptscriptstyle\rm tr}^2$.

To discuss eq.~(\ref{om-fl-meta-eq}) we may now distinguish various
cases depending on whether we have to consider {\it (i)} the
non-stationary or quasi-stationary situation, {\it (ii)} the strong
($\alpha\!\ll\!a/R$) or weak barrier ($\alpha\!\gg\!a/R$) limit, {\it
  (iii)} the formation of the metastable dimer or stable trimer as
rate limiting, {\it (iv)} the encounter processes to be faster or
slower than the escape process ($\omega_n'\tau_n'\!\ll\!1$ or not for
$n\!=\!2,3$), and {\it (v)} $\tau_2'$ to be dominated by the
metastable dimer state ($\tau_2'\!\sim\!\tau_{\rm\scriptscriptstyle
  dis}$ in the weak barrier limit and
$\tau_2'\!\sim\!\tau_1\tau_{\rm\scriptscriptstyle
  dis}/\tau_{\rm\scriptscriptstyle tr}$ in the strong barrier limit)
or to be dominated by the state of separated atoms
($\tau_2'\!\sim\!\tau_2$). Rather than treating all these possible
cases (and analyzing their possible occurrence for the generic growth
law (\ref{R(t)-eq}) by employing self-consistency requirements) we
only remark that the results obtained by Krug {\it et
  al.}\cite{Krug/etal:2000} are entailed in our description. In this
work, certain regimes corresponding to the quasi-stationary case in the
strong barrier limit are considered for both
$q_1\!\sim\!q_3\!\ll\!1$ and $q_1q_2\!\ll\!1$, where we obtain
$p_1\!\simeq\!q_1$ and $p_2\!\simeq\!q_1q_2$ from
eq.~(\ref{pn-meta-eq}). Since
$\omega_1'\tau_1'\!=\!\tau_1/\tau_{\scriptscriptstyle\rm tr}\!\sim
\!a/\alpha R\!\gg\!1$ in the strong barrier limit, we can always set
$p_{\scriptscriptstyle\rm enc}^{\scriptscriptstyle\rm
  (2)}\!\simeq\!1$ in eq.~(\ref{om-fl-meta-eq}). The following
regimes are then discussed in ref.~\onlinecite{Krug/etal:2000} with
increasing $\tau_{\scriptscriptstyle\rm dis}$:
\begin{list}{}{\setlength{\topsep}{0.1cm}\setlength{\leftmargin}{0.7cm}
    \setlength{\labelwidth}{0.6cm}\setlength{\labelsep}{0.2cm}
    \setlength{\itemsep}{0.1cm}\setlength{\parsep}{0cm}}
\item[\it (i)] For $\tau_{\scriptscriptstyle\rm
    dis}\!\ll\!\tau_{\scriptscriptstyle\rm tr}^2/\tau_1$ we have
  $p_{\scriptscriptstyle\rm enc}^{\scriptscriptstyle\rm
    (3)}\!\simeq\!\omega_3'\tau_3'\!\sim\!\tau_{\scriptscriptstyle\rm
    dis}\tau_1/\tau_{\scriptscriptstyle\rm tr}^2$ and
  $\tau_2'\!\sim\!\tau_1$, i.e.\ $p_2\!\simeq\!  q_1^2$ and
  $p_1\!\gg\!p_2p_{\scriptscriptstyle\rm enc}^{\scriptscriptstyle\rm
    (3)}$.  Accordingly, we obtain $\Omega_{\scriptscriptstyle\rm
    fl}\!\sim\!(\pi FR^2\tau_1)^3 \tau_{\scriptscriptstyle\rm
    dis}/\tau_{\scriptscriptstyle\rm tr}^2$ corresponding to eq.~(14)
  (regime I) in ref.~\onlinecite{Krug/etal:2000}.
\item[\it (ii)] For $\tau_{\scriptscriptstyle\rm tr}^2/\tau_1\!\ll\!
  \tau_{\scriptscriptstyle\rm dis}\!\ll\!\tau_{\scriptscriptstyle\rm
    tr}$, we find $p_{\scriptscriptstyle\rm
    enc}^{\scriptscriptstyle\rm (3)}\!\simeq\!1$ and
  $\tau_2'\!\sim\!\tau_1$, i.e.\ $p_2\!\simeq\!q_1^2$ and
  $p_1\!\gg\!p_2p_{\scriptscriptstyle\rm enc}^{\scriptscriptstyle\rm
    (3)}$ as in {\it (i)}, and hence $\Omega_{\scriptscriptstyle\rm
    fl}\!\sim\!(\pi FR^2)^3\tau_1^2$ corresponding to eq.~(15) (regime
  II) in ref.~\onlinecite{Krug/etal:2000}. In the following cases,
  where $\tau_{\scriptscriptstyle\rm dis}$ becomes even larger (and
  $\tau_1$, $\tau_{\scriptscriptstyle\rm tr}$ do not change), we still
  have $p_{\scriptscriptstyle\rm enc}^{\scriptscriptstyle\rm
    (3)}\!\simeq\!1$.
\item[\it (iii)] For $\tau_{\scriptscriptstyle\rm
    tr}\!\ll\!\tau_{\scriptscriptstyle\rm dis}\!\ll\!
  \tau_{\scriptscriptstyle\rm tr}/\pi FR^2\tau_1$,
  $\tau_2'\!\sim\!\tau_1\tau_{\scriptscriptstyle\rm
    dis}/\tau_{\scriptscriptstyle\rm tr}$, i.e.\ 
  $p_2\!\sim\!q_1q_2\!\sim\!(\pi
  FR^2)^2\tau_1^2\tau_{\scriptscriptstyle\rm
    dis}/\tau_{\scriptscriptstyle\rm tr}$. The condition
  $\tau_{\scriptscriptstyle\rm dis}\!\ll\!\tau_{\scriptscriptstyle\rm
    tr}/\pi FR^2\tau_1$ is equivalent to
  $p_1\!\gg\!p_2p_{\scriptscriptstyle\rm enc}^{\scriptscriptstyle\rm
    (3)}\!\simeq\!p_2$, and we thus find
  $\Omega_{\scriptscriptstyle\rm fl}\!\simeq\!\pi
  FR^2p_2\!\simeq\!(\pi FR^2)^3\tau_1^2\tau_{\scriptscriptstyle\rm
    dis}/\tau_{\scriptscriptstyle\rm tr}$ corresponding to eq.~(16)
  (regime III) in ref.~\onlinecite{Krug/etal:2000}.
\item[\it (iv)] For $\tau_{\scriptscriptstyle\rm tr}/\pi
  FR^2\tau_1\!\ll\!\tau_{\scriptscriptstyle\rm dis}$ finally,
  $p_1\!\gg\!p_2$ and the formation of the stable trimer is no longer the
  rate limiting process.  From eq.~(\ref{om-fl-meta-eq}) we then
  obtain $\Omega_{\scriptscriptstyle\rm fl}\!\simeq\!\pi
  FR^2p_1\!\sim\!(\pi FR^2)^2\tau_1$ corresponding to eq.~(17) (regime
  IV) in ref.~\onlinecite{Krug/etal:2000}. As expected, the scaling
  behavior of $\Omega_{\scriptscriptstyle\rm fl}$ in this limit
  reduces to the case $i\!=\!1$ (see also the discussion in
  Sec.~\ref{nonnegligible-subsec}).
\end{list}

It is clear that the above analysis is difficult to extend to even
more complicated situations.  Moreover, due to the growing number of
characteristic time scales, we found it increasingly difficult to
discern pronounced scaling regimes in practice, see e.g.\ 
Fig.~\ref{rc-beta-fig}. We therefore prefer to treat the problem of
second layer nucleation in the presence of metastable clusters within
the more general framework outlined in the following Section.

\section{Second layer nucleation in general situations}
\label{general-sec}
In a more general approach to the problem of second layer nucleation
we distinguish between different individual states of the island
during its growth with respect to the number of atoms that are on top
of the island and the way a given number of atoms is decomposed into
clusters of various sizes. Employing a Poisson approximation, the
transition processes between the states exhibit no (intrinsic) memory
and can be characterized by elementary rates. For the non-interacting
particle model these elementary transition rates are the deposition
rate $\pi FR^2$, the rate for the attachments of a single atom to an
intermediate cluster of size $k$, and the loss rate of adatoms.  The
latter is given by the inverse lifetime $\tau_1^{-1}$ of a single atom
(see eq.~(\ref{taun-eq})). Dissociation rates enter the problem as
additional parameters, when the lifetimes of intermediate metastable
clusters can not be neglected. The consequences of such dissociation
rates will be discussed in Sec.~\ref{nonnegligible-subsec}.  First,
however, we will present the general procedure in
part~\ref{general-proc-subsec} and show in
part~\ref{negligible-subsec} how the results of the simplified
stochastic description in Sec.~\ref{simple-sec} can be recovered. In
addition to these previously derived results, it is also discussed how
the general treatment allows one to gain detailed insight into the
dominant microscopic pathways that are followed to form a stable
nucleus on top of the island.

\subsection{General Procedure}
\label{general-proc-subsec}
Let us introduce a common notation for the elementary transition
rates: $W_{\rm F}$ for the deposition rate, $W_{\rm l}$ for the loss
rate, $W_{{\rm a},j}^{(n)}$ for the attachment rate for a single atom
to an intermediate cluster of size $j$ if in total $n$ atoms are
present on top of the island (see eq.~\ref{ws-all-eq} in
Appendix~\ref{omn-app}; we formally include the case $j\!=\!1$), and
$W_{{\rm d},j}$ for the dissociation rate of an unstable cluster
composed of $j\!\le\!i$ atoms (again we do not distinguish between
different cluster configurations for the same cluster size, see also
the remark in ref.~\onlinecite{nucleus-comm}).  According to the
results derived in Sec.~\ref{simple-sec} and Appendix~\ref{omn-app}
the transition rates are
\begin{mathletters}
\label{wall-eq}
\begin{equation}
W_{\rm F}=\pi FR^2\,,
\label{wf-eq}
\end{equation}
\begin{equation}
W_{\rm l}=
\frac{D}{R^2}\,\left(\kappa_1\frac{a}{\alpha R}+\kappa_2 \right)^{-1}\,,
\label{wl-eq}
\end{equation}
\begin{equation}
W_{{\rm a},j}^{(n)}=\kappa_{\rm a,j}^{(n)}
\left\{\begin{array}{l@{\hspace*{0.5cm}}c}
\displaystyle \frac{n(n\!-\!1)}{2}\,\frac{2D}{\pi R^2}\,, & 
\displaystyle j\!=\!1 \\[0.5cm]
\displaystyle (n\!-\!j)\,\frac{D}{\pi R^2}\,, & 
\displaystyle 2\!\le\!j\!\le\!n
\end{array}\right.
\label{wa-eq}
\end{equation}
\begin{equation}
W_{{\rm d},j}=\kappa_{\rm d,j}\frac{D}{a^2}
 \exp\left(-\frac{\Delta E_j^{\scriptscriptstyle\rm dis}}{k_{\rm B}T}\right)\,,
\hspace*{0.4cm}2\le j\le i\,,
\label{wdj-eq}
\end{equation}
\end{mathletters}
where $\Delta E_j^{\scriptscriptstyle\rm
  dis}\!=\!E_j^{\scriptscriptstyle\rm dis}\!-\!E_0$ is the
dissociation energy of a {\it single} atom from an unstable cluster of
size $j\le i$. The prefactors $\kappa_{\rm a,j}^{(n)}$ and
$\kappa_{\rm d,j}$ contain the effective sizes of cluster perimeters
on one hand (see the discussion in Appendix~\ref{omn-app}), and
various corrections involved in the overall approximation scheme
(Poisson approximation, cutoff $n_\star$ 
%****************************************************************
\begin{figure}[t!]
\begin{center}\epsfig{file=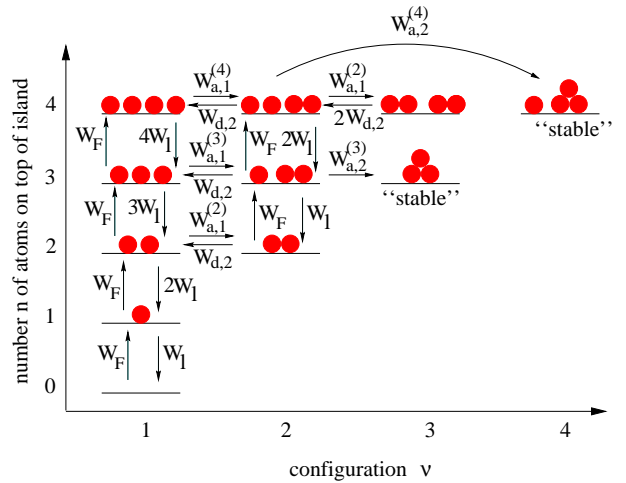,width=8cm}\end{center}
\vspace*{-0.2cm}
\caption{Various states and the corresponding transition rates
  (see eq.~(\ref{wall-eq})) involved in the second layer nucleation
  for a critical nucleus of size $i=2$.}
\label{rate-illus1-fig}
\end{figure}
\vspace*{-0.2cm}\noindent
%****************************************************************
introduced below, etc.); they
are considered to be independent of $D$, $\alpha$ and $R$. In
principle one should also take into account the possibility that a
subcluster composed of more than one atom can dissociate from an
unstable cluster. In fact, it has been argued that such cluster
dissociations are sometimes more likely to occur than the dissociation
of single atoms, as e.g.\ for dimer dissociation from a tetramer on a
(100) surface by a kind of ``shearing mode''.\cite{Shi/etal:1996} For
simplicity we will take into account only the dissociation of single
atoms here, although conceptually the inclusion of cluster
dissociation processes into the general treatment poses no difficulty.
Also, we do not consider the influence of cluster mobilities. If one
would allow for a small jump rate $D_j/a^2$ of a cluster of size
$j\!\ge\!2$, the relative diffusion of a $j$ cluster and a single atom
would be larger by a factor $1\!+\!D_j/D$ and accordingly we had to
multiply $W_{{\rm a},j}^{(n)}$ in eq.~(\ref{wa-eq}) by this factor for
$j\!\ge\!2$.

The method is best introduced by an example. To this end consider
Fig.~\ref{rate-illus1-fig} that illustrates the situation for a
critical nucleus of size $i\!=\!2$. Various states of the island are
shown, which are distinguished according to the total number $n$ of
atoms on top of the island, and the possible configurations that can
be assumed for a given $n$. Between the states the possible
transitions are marked by arrows that are labeled by the
corresponding rates. Note that the loss from a state with $n$ single
atoms is $n$ times larger than the loss from the state with one atom.
It is clear that Fig.~\ref{rate-illus1-fig} shows only a small part of
the possible states and in principle can be extended by including
larger numbers $n$. However, as will be pointed out below, these
states with larger $n$ do not contribute much to the onset of second
layer nucleation. Moreover, we have not included states containing
stable clusters of size $j\!>\!i\!+\!1$ and transitions between
different states containing a stable nucleus of size $i\!+\!1$. These
are irrelevant for the fraction $f_0(t)$ of covered islands at time
$t$.

We denote by $p_{n,\nu}$ the probability for the island to be in state
$(n,\nu)$, where $n$ refers to the number of atoms on top of the
island and $\nu$ to a specific configuration for a given $n$.  A
complete description of the stochastic process amounts to specifying
the set $\{p_{n,\nu}(t)\}$ of state probabilities at all times $t$.
The time evolution of the $\{p_{n,\nu}(t)\}$ is described by the
master equation
\begin{eqnarray}
\frac{dp_{n,\nu}}{dt}&=&\sum_{n',\nu'}
\bigl[W(n',\nu'\!\to\!n,\nu)\,p_{n',\nu'}\nonumber\\
&&\phantom{\sum_{n',\nu'}}
-W(n,\nu\!\to\!n',\nu')\,p_{n,\nu}\bigr]\,,
\label{master-eq}
\end{eqnarray}
where for the rates $W(n,\nu\!\to\!n',\nu')$ the appropriate
expressions from eq.~(\ref{wall-eq}) have to be substituted (see
Fig.~\ref{rate-illus1-fig}). Note that transitions are possible only
between a limited number of states. In the situation considered here,
where only single atoms can leave the island, we have $W(n,\nu\to
n',\nu')\!=\!0$ for $|n-n'|\!\ge\!2$.

To treat the problem of second layer nucleation under generic growth
conditions one has to solve the set of eqs.~(\ref{master-eq}) for
$R\!=\!R(t)$ with $R(t)$ from eq.~(\ref{R(t)-eq}) subject to the
initial condition $p_{n,\nu}\!=\!\delta_{n,0}$.  To this end it is
convenient to solve (\ref{master-eq}) using $R$ as the independent
variable. The integration of the differential equations
(\ref{master-eq}) using standard solvers takes very little CPU time on
ordinary workstations, so that results for $f_0(t)$ and $R_c(\alpha)$
can be obtained almost immediately. Numerical results are discussed in
the following.

\subsection{Negligible Lifetimes of Unstable Clusters}
\label{negligible-subsec}
In this subsection we consider the case $\Delta
E_j^{\scriptscriptstyle\rm dis}\!=\!0$ that was treated extensively in
Sec.~\ref{simple-sec}. The fraction $f_0(t)$ of covered islands
within our more general framework is given by
\begin{equation}
f_0(t)=\sum_{n=i\!+\!1}^\infty p_{n,\nu_n}(t)\,,
\label{f0rate-eq}
\end{equation}
where $\nu_n$ is the configuration containing a stable nucleus for a
given $n$ (for example, $\nu_3\!=\!3$ and $\nu_4\!=\!4$ in
Fig.~\ref{rate-illus1-fig}).  In practice, states corresponding to
large $n$ contribute a negligible amount up to times $t_c$, so that
one
needs to consider a finite maximum number of atoms $n_\star$ only
($n_\star\!=\!4$ in Fig.~\ref{rate-illus1-fig} turned out to be
sufficient).

\begin{figure}[t]
\begin{center}\epsfig{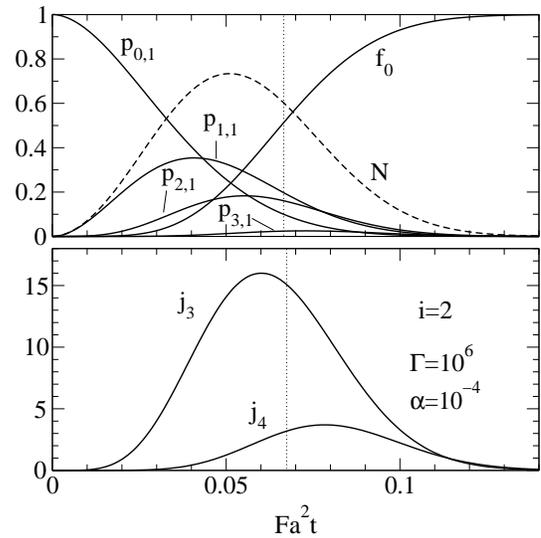}\end{center}
\vspace*{-0.2cm}
\caption{Time development of the fraction of covered islands $f_0(t)$,
  the mean number of atoms $N(t)$ in states containing no stable
  nucleus, the occupation probabilities $p_{n,1}$, and the currents
  $j_3$ and $j_4$ (see eq.~\ref{jn(t)-eq}) for $i\!=\!2$,
  $\alpha=10^{-4}$ and $\Gamma=10^6$. The maximum number of atoms is
  $n_\star\!=\!4$ corresponding to the digram shown in
  Fig.~\ref{rate-illus1-fig}. The vertical dotted line marks the
  coverage $Fa^2t_c$ at the critical time $t_c$.}
\label{states-i2-fig}
\end{figure}

Figure~\ref{states-i2-fig}a shows $f_0(t)$ and the probabilities
$p_{n,\nu}(t)$ (labeled according to Fig.~\ref{rate-illus1-fig}) as a
function of the coverage $Fa^2t$ for $\alpha\!=\!10^{-4}$, and
$\Gamma\!=\!10^6$. Also shown is the mean total number
\begin{equation}
N(t)\equiv\sum_{n=1}^{n_\star}\sum_{\nu\ne\nu_n} p_{n,\nu}(t)\,n
\label{N(t)-eq}
\end{equation}
of atoms that are not in states possessing a stable nucleus. In
accordance with the predictions of the simplified stochastic
description, this number is less than one 
%****************************************************************
\noindent
%****************************************************************
up to time $t_c$.
Accordingly, the pathway followed by the system to form a stable
nucleus is dominated by fluctuations as discussed in
Sec.~\ref{simple-sec}. The important role of the fluctuations can
even more clearly be recognized by looking at the state probabilities
$p_{n,\nu}$, $\nu\!\ne\!\nu_n$, and the currents
\begin{equation}
j_3(t)\equiv
W_{{\rm a},2}^{(3)}\,p_{3,2}(t)\,,\hspace*{0.5cm}
j_4(t)\equiv
W_{{\rm a},2}^{(4)}\,p_{4,2}(t)
\label{jn(t)-eq}
\end{equation}
into the states containing a stable nucleus. As can be seen from
Fig.~\ref{states-i2-fig}a, only the probabilities $p_{n,1}(t)$ are
significant, while the other state probabilities $p_{n,2}(t)$ and
$p_{4,3}(t)$ cannot be discerned on the scale used in the figure. On
the other hand, we find that the current $j_3(t)$ from the state
$(n\!=\!3,\nu\!=\!2)$ (which has a very small probability
$p_{3,2}(t)$) contributes most to the growth of $f_0(t)$, see
Fig.~\ref{states-i2-fig}b. The fact that $j_4(t)$ gives only a subdominant
contribution to second layer nucleation, indicates that the
incorporation of states with $n\!>\!5$ will not significantly change
the behavior of $f_0(t)$.

\begin{figure}[t]
\begin{center}\epsfig{file=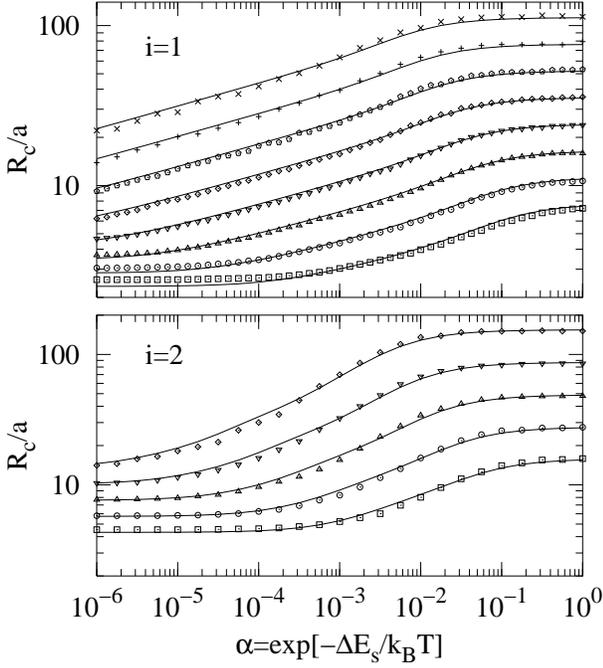,width=8.5cm}\end{center}
\vspace*{-0.2cm}
\caption{Comparison of critical island sizes $R_c$ obtained from the
  rate-equation approach (solid lines) with Monte Carlo data for
  critical nuclei $i=1,2$ and different values of $\Gamma$ (same
  symbols as in Figs.~\ref{rc-i1-fig},\ref{rc-i2-fig}).}
\label{rc-rate-fig}
\end{figure}

The results for $f_0(t)$ compare well with the data obtained from
kinetic Monte Carlo simulations, the quality of agreement between
theory and simulation being as good as in Fig.~\ref{f-t-fig}. The
values of the optimal prefactors $\kappa_{\rm a,j}^{(n)}$ and
$\kappa_{\rm d,2}$ are given in ref.~\onlinecite{prefactor-comm}. To
exemplify the good agreement between theory and simulations, we have
re-plotted in Fig.~\ref{rc-rate-fig} the critical radius $R_c$ as a
function of $\alpha$ for various $\Gamma$ for $i\!=\!1,2$ from
Figs.~\ref{rc-i1-fig},\ref{rc-i2-fig}. The solid lines referring to
the numerical results give an excellent fit to the Monte Carlo data.
For $i\!=\!1$, only the states with $n\!\le\!2$ in
Fig.~\ref{rate-illus1-fig} had to be included to achieve this almost
perfect agreement.

%****************************************************************
\noindent
%****************************************************************
For $i\!>\!2$ we expect that a very large number $n_\star$ has to be
chosen in order to obtain a correct description of second layer
nucleation within the rate equation approach. Diagrams corresponding
to that shown in Fig.~\ref{rate-illus1-fig} then become very
complicated and not easily tractable from the practical point of view.
It is thus helpful to introduce
%****************************************************************
\begin{figure}[b]
\begin{center}\epsfig{file=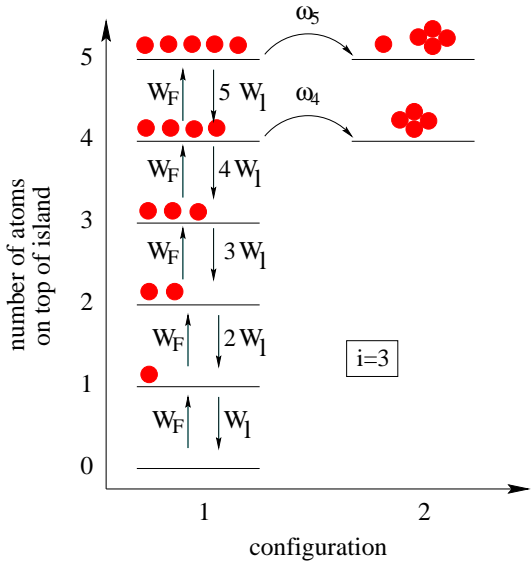,width=7cm}\end{center}
\vspace*{-0.2cm}
\caption{States and corresponding transition rates
  (see eqs.~(\ref{wall-eq},\ref{omn-eq})) involved in the second layer
  nucleation for a critical nucleus of size $i\!=\!3$ when the
  lifetimes of unstable clusters are neglected. Only states
  corresponding to $n\!\le\!5$ atoms on top of the island are shown.}
\label{rate-illus2-fig}
\end{figure}
\vspace*{-0.2cm}\noindent
%****************************************************************
\begin{figure}[t]
\begin{center}\vspace*{-0.5cm}
\epsfig{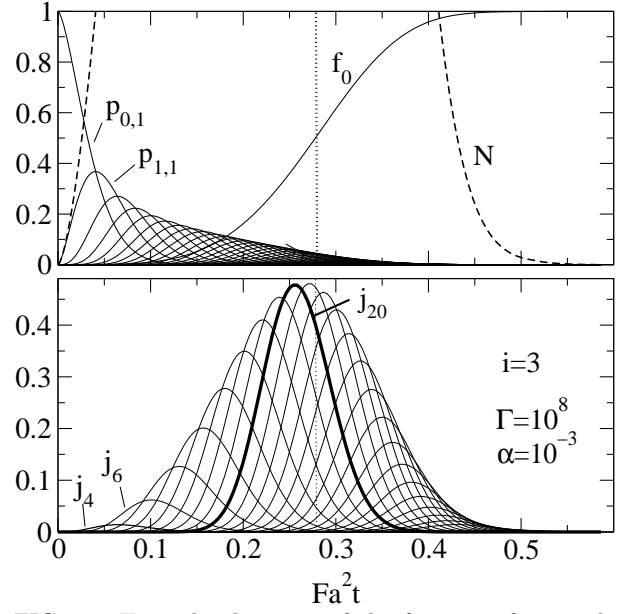}\end{center}
\vspace*{-0.2cm}
\caption{Time development of the fraction of covered islands
  $f_0(t)$, the mean number of atoms $N(t)$ in states containing no
  stable nucleus, the occupation probabilities $p_{n,1}$, and the
  currents $j_n(t)=\omega_np_{n,1}$, $n\!\ge\!i\!+\!1$ for $i\!=\!3$,
  $\alpha=10^{-3}$ and $\Gamma=10^8$. The maximum number of atoms
  included in a diagram of the type shown in
  Fig.~\ref{rate-illus2-fig} is $n_\star\!=\!50$.  The vertical dotted
  line marks the coverage $Fa^2t_c$ at the critical time $t_c$. At
  $t_c$ we find $N(t_c)\!\protect\cong\!10$.}
\label{states-i3-fig}
\end{figure}\vspace*{-0.2cm}\noindent
%****************************************************************
the ``renormalized'' encounter rates
$\omega_n$ defined in eq.~(\ref{omn-eq}) and to consider simplified
diagrams as shown in Fig.~\ref{rate-illus2-fig} for $i\!=\!3$. For a
given number $n\!\ge\!i\!+\!1$ of atoms on top of the island we have
only included two states $\nu\!=\!1,2$: One of these refers to a
state where all $n$ atoms are separated ($\nu\!=\!1$), and the other
one to a state, where exactly $i\!+\!1$ atoms form a stable nucleus,
while the remaining $n\!-\!(i\!+\!1)$ atoms are not bound to other
atoms in the same layer $(\nu\!=\!2)$. 

Plots of $f_0(t)$, $p_{n,1}(t)$, $N(t)$, and
$j_n(t)\!\equiv\!p_{n,1}(t)\omega_n(R(t))$ for $i\!=\!3$ analogous to
Fig.~\ref{states-i2-fig} are shown in Fig.~\ref{states-i3-fig}. As expected
from the discussion in Sec.~\ref{simple-sec}, we now had to take into
account states with $n$ up to $n_\star\!=\!50\!\gg\!i(i+1)/2\!=\!6$
before reaching the limit, where $f_0(t)$ as calculated from
eq.~(\ref{f0rate-eq}) did not change much by incorporation of states
with larger $n$. Near $t_c$, $N(t)$ is significantly larger than
$i\!+\!1\!=\!4$ (at $t_c$ we find $N(t_c)\cong10$), and the dominant
currents $j_n(t)$ initiating second layer nucleation are those for
$n\!=\!15-20\!\gg\!4$, see Fig.~\ref{states-i3-fig}.

\begin{figure}[t]
\begin{center}\epsfig{file=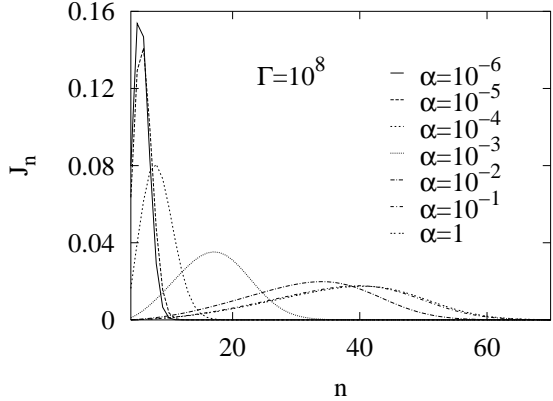,width=8cm}\end{center}
\vspace*{-0.2cm}
\caption{Integrated nucleation currents up to time $t_c$ as a function
  of the number of particles $n$ in the state from which the
  nucleation event took place ($i\!=\!3$).}
\label{Jn-fig}
\end{figure}

In order to see how the preferred paths for second layer nucleation
change with the step crossing probability $\alpha$, we define the
integrated current $j_n(t)$ up to $t_c$ by
\begin{equation}
J_n\!\equiv\!\int_0^{t_c}\hspace*{-0.25cm} dt\,j_n(t)\!=\!\int_0^{t_c}
\hspace*{-0.15cm} dt\,
p_{n,1}(t)\omega_n(R(t))\,,\hspace*{0.1cm}n\ge i\!+\!1\,.
\label{Jn-eq}
\end{equation}
This quantity equals the fraction of covered islands at time $t_c$ for
which the stable nucleus originates from 
a state possessing exactly
$n$ adatoms. Figure~\ref{Jn-fig} shows $J_n$ as a function of $n$ for
fixed $\Gamma=10^8$ and various $\alpha$. We
%****************************************************************
\vspace*{-0.2cm}\noindent
%****************************************************************
see that for all $\alpha$
states with $n\!\gg\!4$ dominate the onset 
of the nucleation. The
number of particles $n_{\rm peak}$ in the state where $J_n$ has a
maximum strongly increases with increasing $\alpha$. For
$\alpha\!=\!10^{-3}$, the second layer nucleation is typically
initiated by $n_{\rm peak}\!\simeq\!18$ adatoms on top of the island.

\subsection{Influence of Metastable Clusters}
\label{nonnegligible-subsec}
The general procedure outlined in Sec.~\ref{general-proc-subsec} allows us
also to describe situations, where the binding energies of unstable
clusters of size $j\!\le\!i$ are not small compared to $k_{\rm
  B}T$. To demonstrate this we again consider the case $i\!=\!2$ and
the corresponding diagram in Fig.~\ref{rate-illus1-fig}. The dimer in
the intermediate states possessing no stable nucleus is now considered
to be metastable, and we introduce the parameter
\begin{equation}
\beta\equiv\exp(-\Delta E_j^{\scriptscriptstyle\rm dis}/k_{\rm B}T)
\label{beta-eq}
\end{equation}
as ``dissociation probability'' (analogous to the step edge crossing
probability $\alpha=\exp(-\Delta E_{\rm S}/k_{\rm B}T)$). For $\beta=1$
we recover the non-interacting particle model.  From the outset it is
clear that second layer nucleation will proceed faster for smaller
$\beta$, since the state probabilities $p_{3,2}(t)$ 
and $p_{4,2}(t)$
in Fig.~\ref{rate-illus1-fig} and accordingly the currents $j_3(t)$
and $j_4(t)$ defined in eq.~\ref{jn(t)-eq} will become strongly
enhanced.

\begin{figure}[b]
\begin{center}
\vspace*{-0.7cm}\epsfig{file=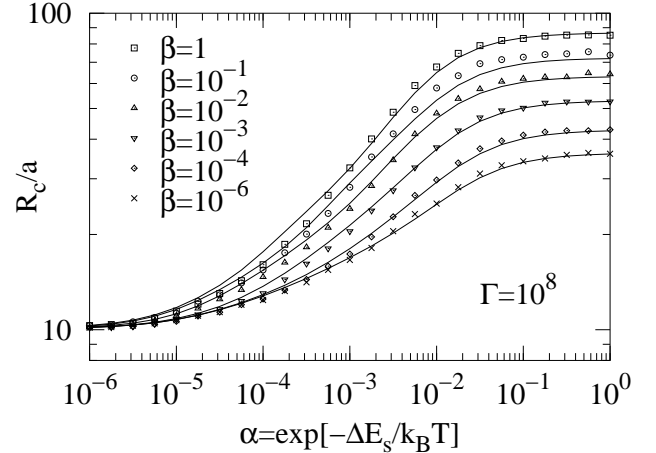,width=9cm}\end{center}
\vspace*{0cm}
\caption{Comparison of critical island sizes $R_c$ obtained from the
  rate-equation approach with results from kinetic Monte Carlo
  simulations of the one-island model for $i\!=\!2$. Dimers are
  metastable and dissociate with a rate $\beta D/a^2$ (see
  eq.~(\ref{beta-eq})).}
\label{rc-beta-fig}
\end{figure}

Figure~\ref{rc-beta-fig} shows $R_c$ as a function of $\alpha$ for
fixed $\Gamma\!=\!10^8$ and various $\beta$ obtained from Monte Carlo
simulations (open symbols). As expected, the critical radius decreases
with decreasing $\beta$. In fact, for $\beta\!=\!10^{-6}$ one can
regard the dimer as effectively being stable on the relevant time
scale $t_c$, so that the changes with $\beta$ correspond to a
continuous transition from $i\!=\!2$ ($\beta\!=\!1$) to $i\!=\!1$
($\beta\!=\!10^{-6}$). The comparison with the numerical solution of
eqs.~\ref{master-eq} (solid lines) yields very good agreement. To
achieve this agreement, we used the {\it same} set of prefactors
$\kappa_{\rm a,j}^{(n)}$, $\kappa_{\rm d,2}$ as in
Fig.~\ref{rc-rate-fig}.\cite{prefactor-comm} This highlights the power
of the rate equation approach to treat second layer nucleation in
general situations.
%****************************************************************
%****************************************************************

\section{Summary and Discussion}
\label{summary-sec}
In summary, we have presented a detailed theoretical investigation of
the nucleation on top of islands in epitaxial growth. In the
non-interacting particle model, where the lifetimes of unstable
clusters can be neglected, it was possible to tackle the problem
within a simplified stochastic description based on scaling arguments.
An important result for the non-interacting case is that the nucleation
for critical nuclei of size $i\!\le\!2$ is dominated by fluctuations,
while for larger critical nuclei it can be treated in a mean-field type
manner. The second layer nucleation rate for both cases was derived in
compact form (see eqs.~(\ref{om-fl-eq},\ref{om-mf-eq})). When metastable
clusters can form with appreciable lifetimes, the simplified
description can in principle be extended (see
Sec.~\ref{simple-meta-subsec}), but becomes of limited value due to
the fact that many elementary processes get mutually coupled both
sequentially and in parallel. In such situations it is better to
employ the more general framework outlined in Sec.~\ref{general-sec}
that is based on our derivation for the transition rates of the
elementary processes. Results obtained from both theoretical approaches
were shown to agree with Monte Carlo data.

Throughout the paper, we have used the generic growth law
(\ref{R(t)-eq}) for the mean island radius, but it is straightforward
to treat other growth laws also (as e.g.\ an exponential behavior),
which may be realized by special preparation
techniques.\cite{Bromann/etal:1995} Neither the general expressions
(\ref{om-mf-eq},\ref{om-fl-eq}) for the second layer nucleation rates
in simple situations nor the master equation (\ref{master-eq}) depend
on the specific form of the growth law (the expressions for $\bar n$,
$p_n$, etc.\ in the quasi-stationary case however get modified, see
the discussion in Sec.~\ref{simple-proc-subsec}). Moreover, it is
straightforward to rewrite all formulae for the case of heteroepitaxy
by replacing the jump rate $D/a^2$ of adatoms on top of the islands by
a modified jump rate $D'/a^2$.

The theoretical understanding of second layer nucleation is not only
of basic importance but has numerous applications. One of these is the
determination of the effective step edge barrier $\Delta E_{\rm S}$
for systems, where the more direct and simpler method via the
measurement of adatom lifetimes by field ion
microscopy\cite{Kyuno/Ehrlich:1998} cannot be applied. As pointed out
in ref.~\onlinecite{Rottler/Maass:1999}, the breakdown of the TDT
approach in the fluctuation-dominated situation calls for a
reexamination of some experimental data for estimating $\Delta E_{\rm
  S}$.  In fact, such reexamination has been carried out recently by
Krug {\it et al.}\cite{Krug/etal:2000} with notable results: By
reanalyzing the fraction of covered islands $f(t)$ measured for
Ag/Ag(111)\cite{Bromann/etal:1995} they corrected the previously
reported estimate $\Delta E_s\!\approx\!0.12{\rm eV}$ to $\Delta
E_s\!\approx\!0.32{\rm eV}$ (they also reported another estimate
yielding $\Delta E_s\!\approx\!0.20{\rm eV}$ based on a modified data
analysis, see the comment in ref.~\onlinecite{ag-comm}). Krug {\it et
  al.}  moreover studied the influence of step decoration by CO
molecules\cite{Kalff/etal:1998} on $R_c$ (and hence $\Delta E_{\rm
  S}$) for Pt/Pt(111). They found a strong increase of $\Delta E_{\rm
  S}$ with CO partial pressures, when analyzing the data corresponding
to regime II (for $i\!=\!1$) of the fluctuation-dominated situation.
Hence contamination by CO is expected to favor multilayer growth.

On the other hand, surfactants may promote smooth layer-by-layer
growth. For example, the presence of only small amounts of Sb for
growth of Ag on Ag(111) were shown to convert rough multilayer to
layer-by-layer growth.\cite{Vegt/etal:1992,Vrijmoeth/etal:1994} It was
suggested\cite{Vegt/etal:1992} that Sb reduces $\Delta E_{\rm S}$,
but, since it was observed that Sb increases the island density in the
first layer,\cite{Vegt/etal:1992,Tersoff/etal:1994} it is also
possible that the induced layer-by-layer growth results from a
decrease of the mean island distance. Even in the absence of
surfactants, a change of the {\it effective} step edge barrier may go
along with a shape transition of the islands with varying temperature
(see
refs.~\onlinecite{Kunkel:1990,Bott/etal:1992,Smilauer:1993,Tersoff/etal:1994}
and the comment in ref.~\onlinecite{effes-comm}), and this can induce
changes in the film morphology as well. With respect to the transition
from the fluctuation-dominated to the mean-field type situation with
varying $i$ predicted in this work, it would also be interesting to
conduct proper experiments for metal epitaxy on (100) surfaces, where
a change from $i\!=\!1$ to $i\!=\!3$ is often observed with increasing
temperature.

A further application pertaining to the design of self-organized
nanostructures is the possibility to create pyramidal mounds on a
substrate, which are called ``wedding
cakes''.\cite{Ernst/etal:1994,Kalff/etal:1999,Ernst/etal:2000} As
suggested by Michely {\it et al.},\cite{Michely/etal:1999} the size
$L_{\rm top}$ of the top terrace of the pyramid should be roughly
given by $\Omega(L_{\rm top})\approx F$, where $\Omega$ is the second
layer nucleation rate.  Recently, an expression for the distribution
of $L_{\rm top}$ has been suggested within a self-consistent analysis
of a model for the dynamics of the top terrace.\cite{Krug/etal:2000}
In recent developments of nanostructure formation also larger clusters
of atoms are considered as basic building blocks in epitaxial growth.
The underlying processes seem to be very similar to the case of
deposition of single atoms or simple molecules (for a recent review,
see Ref. \onlinecite{Jensen:1999}), so that it could well be that also for
cluster deposition an effective step edge barrier has a decisive
influence on the film topography.

In light of the basic importance and the manifold applications, there
is certainly need for further improvement of our understanding of
second layer nucleation. Topics worthy of further study are in
particular the influence of strain effects and longer-range
interactions between the adatoms. The latter may be attributed to
direct forces (e.g.\ induced dipole-dipole forces in the case of
magnetic adsorbates), or they can be mediated by perturbations of the
electron structure of the substrate. By extending the approach
presented in this work, these issues may be tackled in the near
future.

\acknowledgments We thank R.~J.~Behm, H.~Brune, W.~Dieterich, and
H.~J.~Ernst for very interesting discussions. P.M. thanks the
Deutsche Forschungsgemeinschaft for financial support (SFB~513, Ma
1636/2).

\appendix

\section{}
\label{omn-app}
\begin{figure}[t!]
\begin{center}\epsfig{file=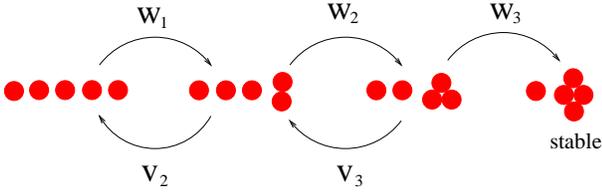,width=8cm}\end{center}
\vspace*{-0.2cm}
\caption{Illustration of the encounter of 4 atoms out of $n\!=\!5$ adatoms.
  The 4 connected atoms are supposed to form a stable cluster
  ($i\!=\!3$).  First a dimer forms with a rate $w_1$ then a trimer
  with a rate $w_2$, and finally the stable quadrumer with a rate
  $w_3$.  The sequential process only leads to the formation of a
  stable cluster, if neither the dimer dissociates with the rate $v_2$
  nor the trimer dissociates with the rate $v_3$.}
\label{omn-illus-fig}
\end{figure}

We want to calculate the characteristic rate $\omega_n(R)$ for an
encounter of $i\!+\!1$ atoms, if initially $n\!\ge\!i\!+\!1$ atoms are
randomly placed on top of an island with radius $R$
and infinite step
edge barrier ($\alpha\!=\!0$). For this purpose let us consider the
encounter as a sequential process as depicted in
Fig.~\ref{omn-illus-fig} (for $i\!=\!3$ and $n\!=\!5$): First a dimer
forms, then one of the remaining atoms attaches to the dimer and a
trimer is created, and so on until a stable cluster composed of
$i\!+\!1$ atoms has been formed. Denoting the rate for the formation of the
dimer by $w_1$, and the rate for the attachment of an atom to an
already existing cluster composed of $k$ atoms (``$k$-cluster'') by
$w_k$, we may write
\begin{mathletters}
\label{ws-all-eq}
\begin{equation}
w_1=\frac{n(n\!-\!1)}{2}\,\frac{2D}{a^2}\,\frac{b_1 a^2}{\pi R^2}\,,
\label{w1-eq}
\end{equation}
\begin{equation}
w_k=(n\!-\!k)\,\frac{D}{a^2}\,\frac{b_k a^2}{\pi R^2}\,,\hspace*{0.6cm}
2\!\le\!k\!\le\!i\,.
\label{ws-eq}
\end{equation}
\end{mathletters}
The factors $b_k$ can be viewed as the effective number of perimeter
sites of a $k$-cluster. Similarly, we may write\linebreak 
for the rate of dissociation $v_k$ of a single atom from a $k$-cluster
(in the case of negligible binding energies of unstable clusters)
\begin{equation}
v_k=d_k\,\frac{D}{a^2}\,,\hspace*{0.6cm}
k\!\ge\!2\,,
\label{vs-eq}
\end{equation}
%****************************************************************
\noindent
%****************************************************************
where again $d_k$ has the meaning of an effective number of perimeter
sites. (In principle one may also take into account the possibility
that a subcluster composed of more than one atom can dissociate from
an unstable cluster and other states with various intermediate
unstable clusters of size $2\!\le\!k\!\le\!i$.)

The idea now is to renormalize the process depicted in
Fig.\ref{omn-illus-fig} by replacing it by an effective transition
rate $w_{\rm eff}$ between the initial state composed of $n$ isolated
atoms and the final state containing the stable cluster. Clearly, such
a replacement is only approximately valid. After the replacement, the
encounter rate $\omega_n(R)$ in eq.~(\ref{omn-eq}) can be identified
with $w_{\rm eff}$. In order to derive $w_{\rm eff}$, we consider a
stationary situation, where the probability $p_1$ of the initial state
is kept fixed and a constant current $J$ flows between neighboring
states containing a $k$- and $k\!+\!1$-cluster. We thus write
\begin{equation}
J=w_kp_k-v_{k+1}p_{k+1}\,,\hspace*{0.5cm} 1\!\le\!k\!\le\!i\!-\!1\,,
\label{J1-eq}
\end{equation}
where $p_k$ denotes the probability of the state containing
a $k$-cluster. The set of equations (\ref{J1-eq}) can be readily
solved for $p_i$ yielding
\begin{equation}
J=w_ip_i=w_1p_1\prod_{k=2}^i\frac{w_k}{v_k}-J\sum_{k=2}^i
\prod_{j=k}^i\frac{w_j}{v_j}\,.
\label{J2-eq}
\end{equation}
On the other hand we have
\begin{equation}
J=w_{\rm eff}p_1\,.
\label{J3-eq}
\end{equation}
Eliminating $J$ from eqs.~(\ref{J2-eq},\ref{J3-eq}), we obtain
\begin{equation}
w_{\rm eff}=\frac{w_1\prod_{k=2}^i w_k/v_k}{1+\sum_{k=2}^i
\prod_{j=k}^i w_j/v_j}\,.
\label{weff-eq}
\end{equation}
For large radii $R\!\gg\!a$, it holds $w_j/v_j\!\ll\!1$ so that we can
neglect the sum over $k$ in the denominator on the right hand side of
(\ref{weff-eq}). Hence we find
\begin{equation}
\omega_n(R)\simeq w_{\rm eff}\simeq
\kappa_{\rm e}\left[\prod_{k=0}^i(n\!-\!k)\right]\,\frac{D}{a^2}\,
\left(\frac{a^2}{\pi R^2}\right)^i\,,
\label{omn-app-eq}
\end{equation}
where $\kappa_{\rm e}\!=\!b_1\prod_{k=2}^i b_k/d_k$.

\section{}
\label{taun-app}
The solution of the diffusion problem\cite{continuum-comm}
\begin{mathletters}
\label{diffprob-eq}
\begin{equation}
\frac{\partial\rho}{\partial t}=D\,\Delta\rho\,,
\label{diffproba-eq}
\end{equation}\vspace*{-0.5cm}
\begin{equation}
\frac{\partial\rho}{\partial r}\Bigl|_{r=0}=0\,,\hspace*{0.4cm}
\left[\frac{\partial\rho}{\partial r}+
    \frac{\alpha}{a}\rho\right]_{r=R}=0
\label{diffprobb-eq}
\end{equation}
\end{mathletters}
with the initial condition $\rho({\bf r},t\!=\!0)=1/(\pi R^2)$
has been derived by Harris:\cite{Harris:1995}
\begin{equation}
\rho({\bf r},t)=\sum_{k=1}^\infty\frac{c_k\lambda_k^2}{2\pi R^2
{\textstyle \frac{\textstyle\alpha R}{\textstyle a}}}
\,\frac{J_0({\textstyle\frac{\textstyle\lambda_k r}{\textstyle R}})}
       {J_0(\lambda_k)}\,
\exp\Bigl(-\lambda_k^2\frac{D}{R^2}t\Bigr)\,.
\label{harris-eq}
\end{equation}
Here $J_\nu(.)$ is the Bessel function of $\nu$th order,
$c_k\!\equiv\!4(\alpha R/a)^2/(\lambda_k^2[\lambda_k^2\!+\!(\alpha R/a)^2])$
and $\lambda_k$ is the $k$th root ($\lambda_1\!<\!\lambda_2\!<\!\ldots$)
of 
\begin{equation}
\Bigl(\frac{\alpha R}{a}\Bigr) J_0(\lambda)=\lambda J_1(\lambda)\,.
\label{kappa-eq}
\end{equation}
The solution (\ref{harris-eq},\ref{kappa-eq}) describes the
probability density for a single diffusing atom that at time $t\!=\!0$
is randomly deposited on top of a circular island with a partially
reflecting boundary.  The probability that the atom has not escaped
from the island up to time $\tau$ is $\Psi(\tau)\!=\!2\pi\int_0^R
dr\,r\,\rho({\bf r},t)$, which yields\cite{Harris:1995}
\begin{equation}
\Psi(\tau)=\sum_{k=1}^\infty c_k\exp\Bigl(-\lambda_k^2\frac{D}{R^2}\tau\Bigr)
\label{Psi-eq}
\end{equation}
Note that, since $\Psi(0)\!=\!1$, it must hold $\sum_{k=1}^{\infty} c_k\!=\!1$.

The probability that none of $n$ independent atoms has escaped
from the island up to time $\tau$ is $\Psi(\tau)^n$.
Accordingly, the probability $\phi(\tau)d\tau$ that the {\it first}
atom leaves the island in the time interval $[\tau,\tau\!+\!d\tau]$ is
\begin{eqnarray}
\phi(\tau)&=&-\frac{d\Psi(\tau)^n}{d\tau}
=-n\Psi(\tau)^{n\!-\!1}\,\frac{d\Psi(\tau)}{d\tau}\nonumber\\
&=&n\frac{D}{R^2}\hspace*{-0.15cm}
\sum_{j_1,\ldots,j_n=1}^\infty\hspace*{-0.3cm}
c_{j_1}{\scriptstyle\ldots} c_{j_n}\,\times\label{phi-eq}\\
&&\hspace*{1.8cm}
\times\,\lambda_{j_1}^2
\exp\Bigl[-(\lambda_{j_1}^2\!+\!{\scriptstyle\ldots}\!+\!
\lambda_{j_n}^2)\frac{D}{R^2}\,\tau\Bigr]\,,\nonumber
\end{eqnarray}
from which for the average time
$\tau_n(R)\!\equiv\!\int_0^\infty d\tau\phi(\tau)\tau$ follows:
\begin{equation}
\tau_n(R)=n\frac{R^2}{D}
\sum_{j_1,\ldots,j_n=1}^\infty\hspace*{-0.5cm}
c_{j_1}{\scriptstyle\ldots} c_{j_n}
\frac{\lambda_{j_1}^2}
   {(\lambda_{j_1}^2\!+\!{\scriptstyle\ldots}\!+\!\lambda_{j_n}^2)^2}\,.
\label{bartauexact-eq}
\end{equation}

It is easy to show that $j_{1,k}\!<\!\lambda_k\!<\!j_{0,k}$, where
$j_{\nu,k}$ is the $k$th zero of $J_\nu(.)$. Since $j_{\nu,k}\!\sim\!
(k\!+\!\nu/2\!-\!1/4)\pi$ for $k\!\gg\!\nu$, the terms in the series
of (\ref{bartauexact-eq}) rapidly decrease with increasing $j_k$,
$k\!=\!1,\ldots,n$ (note that $c_j$ depends on $\lambda_j$).
The leading term can be obtained by setting $c_j=\delta_{j,1}$ in
eq.~(\ref{Psi-eq}), which amounts to a Poisson approximation of the
escape process, $\Psi(\tau)\simeq\exp(-\lambda_1 D\tau/R^2)$. Within
this approximation we obtain
\begin{equation}
\tau_n(R)=\frac{1}{n}\frac{R^2}{D}\frac{1}{\lambda_1^2}\,,
\label{bartauapprox-eq}
\end{equation}
where $\lambda_1$ follows from eq.~(\ref{kappa-eq}). In the limit of
small $\alpha R/a\ll1$ one finds $\lambda_1^2\simeq 2\alpha R/a$,
while in the limit of large $\alpha R/a\gg1$, $\lambda_1^2\simeq
j_{0,1}^2$. Combining these two limits yields the interpolation
formula
\begin{equation}
\tau_n(R)\simeq \frac{1}{n}\,
\frac{R^2}{D}\,\left(\kappa_1\frac{a}{\alpha R}+
\kappa_2\right)
\label{bartauinterpol-eq}
\end{equation}
with $\kappa_1\!=\!1/2$ and $\kappa_2\!=\!1/j_{0,1}^2\!\cong\!0.173$.

Knowing $\tau_n(R)$ we can set up the master equation for the
probabilities $p_n(t)$ to find exactly $n$ atoms on top of the island
at time $t$ in the presence of an incoming flux $F$, see
eq.~(\ref{pnmaster-eq}). Introducing the generating function
$Q(z,t)\!\equiv\!\sum_{n=0}^\infty p_n(t)z^n$ we obtain
from eq.~(\ref{pnmaster-eq})
\begin{equation}
\frac{\partial Q}{\partial t}\!=\!
(z\!-\!1)\left[\pi FR^2Q-\frac{1}{\tau_1(R)}
\frac{\partial Q}{\partial z}\right]\,,
\label{Q-eq}
\end{equation}
where $R\!=\!R(t)$ from eq.~(\ref{R(t)-eq}). Transforming variables from
$t$ to $R$ and defining $\tilde Q(z,R)$ by $\tilde
Q(z,R\!=\!R(t))\!\equiv\!Q(z,t)$, eq.~(\ref{Q-eq}) gives ($a\equiv1$ here)
\begin{equation}
\frac{\partial\tilde Q}{\partial R}+
\frac{\varphi\tilde\alpha}{1\!+\!\tilde\alpha R}(z\!-\!1)
\frac{\partial\tilde Q}{\partial z}=
\frac{2\pi}{A^2\Gamma^{i/(i\!+\!2)}}R^3(z\!-\!1)\tilde Q\,,
\label{tildeQ-eq}
\end{equation}
where $\tilde\alpha\!=\!\kappa_2\alpha/\kappa_1$ and
$\varphi\!=\!2A^{-2}\Gamma^{2/(i\!+\!2)}/\kappa_2$
Equation~(\ref{tildeQ-eq}) is a semi-linear partial differential
equation of first order that can be solved by the method of
characteristics. For the initial condition $\tilde Q(z,R=0)\!=\!1$ we
obtain
\begin{equation}
\tilde Q(z,R)=\exp\left[-(1-z)\bar n(R)\right]\,,
\end{equation}
which for $p_n(R)\!=\!p_n(R(t))\!=\![\partial_z^n\tilde
Q(z,R)/n!]_{z=0}$ yields the Poisson distribution (\ref{pn-eq})
with
\begin{eqnarray}
\bar n(R)&=&\frac{2\pi}{A^2\Gamma^{\scriptstyle\frac{i}{i\!+\!2}}}
(1\!+\!\tilde\alpha R)^{-\varphi}
       \int_0^R dx\,x^3\,(1\!+\!\tilde\alpha x)^\varphi
\nonumber\\
&=&\frac{2\pi}{A^2\Gamma^{\scriptstyle\frac{i}{i\!+\!2}}\tilde\alpha^4}
(1\!+\!\tilde\alpha R)^{-\varphi}
\label{barn-app-eq}\\[0.2cm]
&&\hspace*{0.2cm}\times\Biggl[
\frac{(1\!+\!\tilde\alpha R)^{\varphi\!+\!4}-1}{\varphi\!+\!4}
-3\frac{(1\!+\!\tilde\alpha R)^{\varphi\!+\!3}-1}{\varphi\!+\!3}
\nonumber\\
&&\hspace*{0.2cm}\phantom{\times\Biggl[}{}
+3\frac{(1\!+\!\tilde\alpha R)^{\varphi\!+\!2}-1}{\varphi\!+\!2}
-\frac{(1\!+\!\tilde\alpha R)^{\varphi\!+\!1}-1}{\varphi\!+\!1}\Biggr]\,.
\nonumber
\end{eqnarray}
In the quasi-stationary case ($\partial_t Q\!=\!0$ in
eq.~(\ref{Q-eq})) one obtains
%****************************************************************
\begin{figure}[t]
\hspace*{2.5cm}
\epsfig{file=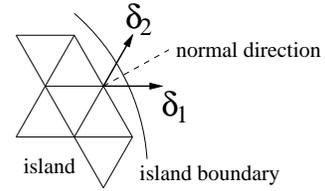,width=4cm}
\vspace*{0.3cm}
\caption{Sketch of the island geometry in the neighborhood
  of a boundary site $\mathbf l$, from which two jumps in the
  directions $\mbox{\boldmath$\delta$}_1$ and
  $\mbox{\boldmath$\delta$}_2$ lead to an escape from the island over
  the step edge barrier.}
\label{boundary-illus-fig}
\end{figure}
\vspace*{-0.2cm}
%****************************************************************
\begin{equation}
\bar n^{\scriptscriptstyle\rm st}(R)=\pi FR^2\tau_1(R)=\pi\frac{FR^4}{D}
\left(\kappa_1\frac{a}{\alpha R}+\kappa_2\right)\,,
\label{barn-st-eq}
\end{equation}
to which (\ref{barn-app-eq}) simplifies for $\varphi\tilde\alpha
R\!\ll\!1$ (see the discussion right after eq.~(\ref{barn-eq})).

Alternatively, we can determine $\bar n^{\scriptscriptstyle\rm st}(R)$
by integrating $\rho_1^{\scriptscriptstyle\rm st}(r)$
over the island area, which yields
\begin{equation}
\bar n^{\scriptscriptstyle\rm st}(R)\!=\!
2\pi\int_0^R\hspace*{-0.2cm}dr\, r\, \rho_1^{\scriptscriptstyle\rm st}(r)\!=\!
\pi \frac{FR^4}{D}\left(\frac{1}{2}\frac{a}{\alpha R}+
\frac{1}{8}\right)\,.
\label{barn-fromrho-eq}
\end{equation}
Hence we can improve the Poisson approximation by renormalizing
the bare coefficient $\kappa_2\!=\!1/j_{0,1}^2\!\simeq\!0.173$ to
$\kappa_2\!=\!1/8$ (note that $\kappa_1\!=\!1/2$ does not change).

Finally, in order to obtain the constants $\kappa_1$ and $\kappa_2$ in
eq.~(\ref{taun-eq}), one has to take into account the ``lattice
corrections''. Let us denote by ${\mathbf l}$ the position of a lattice
site and by $\mbox{\boldmath$\delta$}_j$ the nearest-neighbor bond
vectors, i.e.\ for a triangular lattice
$\mbox{\boldmath$\delta$}_j\!=\!(\cos[2\pi j/6],\sin[2\pi j/6])a$,
$j\!=\!0,\ldots,5$. The master equation describing the diffusion of a
single adatoms on the island reads
\begin{equation}
\frac{\partial w({\mathbf l},t)}{\partial t}=
\frac{D}{6a^2}\sum_{j=0}^5\left[w({\mathbf l}+\mbox{\boldmath$\delta$}_j,t)-
w({\mathbf l},t)\right]\,,
\label{w-master-eq}
\end{equation}
where $w({\bf l},t)$ is the probability to find the atom at lattice
site ${\bf l}$. Equation~(\ref{w-master-eq}) is valid as long as
${\mathbf l}$ is not a boundary site. In the continuum limit we can
write $\sum_{j=0}^5[w({\mathbf l}+\mbox{\boldmath$\delta$}_j,t)-
w({\mathbf l},t)]\!=\!(1/2)\sum_{j=0}^5
(\mbox{\boldmath$\delta$}_j\cdot \mbox{\boldmath$\nabla$})^2w({\mathbf
  l},t)+{\mathcal O}(a^4)$, which yields eq.~(\ref{diffproba-eq}) when
$D$ is replaced by 
$D_{\rm cont}=D/4$ (see also the remark in
ref.~\onlinecite{continuum-comm}). As a consequence, one has to
substitute $D$ by $D_{\rm cont}$ in all continuum equations, in
particular in eq.~(\ref{barn-fromrho-eq}), which means that in
eq.~(\ref{taun-eq}) (referring to the lattice simulations) one should
take $\kappa_1\!=\!4(1/2)\!=\!2$ and $\kappa_2\!=\!4(1/8)\!=\!1/2$.

The value of $\kappa_1$ is still not correct, since we have not taken
into account the lattice correction to the parameter $\alpha$. To
derive this correction we consider a lattice site ${\mathbf l}$ at the
boundary. For example, one may encounter the situation sketched in
Fig.~\ref{boundary-illus-fig}, where $\mbox{\boldmath$\delta$}_0$ and
$\mbox{\boldmath$\delta$}_1$ lead to sites outside the island (where
$w({\mathbf l}+\mbox{\boldmath$\delta$}_j,t)\!\equiv\!0$) and the
remaining nearest neighbor sites ${\mathbf
  l}+\mbox{\boldmath$\delta$}_j$, $j\!=\!2,\ldots,5$ are on the
island. The equation corresponding to (\ref{w-master-eq}) then reads
\begin{eqnarray}
\frac{\partial w({\mathbf l},t)}{\partial t}&=&
\frac{\alpha D}{6a^2}\sum_{j=0}^1
     \left[w({\mathbf l}+\mbox{\boldmath$\delta$}_j,t)-
                    w({\mathbf l},t)\right]\nonumber\\
&&{}
+\frac{D}{6a^2}\sum_{j=2}^5\left[w({\mathbf l}+\mbox{\boldmath$\delta$}_j,t)-
w({\mathbf l},t)\right]\nonumber\\
&=&-\frac{2\alpha D}{6a^2}w({\mathbf l},t)+
\frac{D}{6a^2}\sum_{j=0}^5\left[w({\mathbf l}+\mbox{\boldmath$\delta$}_j,t)-
w({\mathbf l},t)\right]\nonumber\\
&&{}-\frac{D}{6a^2}\sum_{j=0}^1
\left[w({\mathbf l}+\mbox{\boldmath$\delta$}_j,t)-
w({\mathbf l},t)\right].
\label{w-bound-master-eq}
\end{eqnarray}
In a discretization of the second boundary condition in
eq.~(\ref{diffprobb-eq}) on a triangular lattice one has to eliminate
the outer boundary points ${\mathbf l}+\mbox{\boldmath$\delta$}_0$ and
${\mathbf l}+\mbox{\boldmath$\delta$}_1$ via the discretized version
of the ``bulk equation'' (\ref{diffprobb-eq}). This amounts to a
cancellation of the term on the left hand side and the second term on
the right hand side of (\ref{w-bound-master-eq}) in the continuum
limit, and the replacement $\sum_{j=0}^1[w({\mathbf
  l}\!+\!\mbox{\boldmath$\delta$}_j,t)\!-\!  w({\mathbf
  l},t)]\!=\!\sum_{j=0}^1 (\mbox{\boldmath$\delta$}_j\cdot
\mbox{\boldmath$\nabla$})w({\mathbf l},t)+{\mathcal O}(a^2)\!\to\!
\sqrt{3}a\partial w/\partial r$. Hence eq.~(\ref{w-bound-master-eq})
corresponds to the second boundary condition in
eq.~(\ref{diffprobb-eq}), when $\alpha$ is replaced by $\alpha_{\rm
  cont}=2\alpha/\sqrt{3}$ in eq.~(\ref{diffprobb-eq}).

In general, $k$ nearest neighbor sites of a boundary site ${\mathbf
  l}$ can lie outside the island ($k\!=\!1,\ldots,4$). The weights how
often such ${\mathbf l}$ occur and the way the normal direction is
oriented with respect to the nearest neighbor bond vectors leading to
the sites outside the island depend sensitively on the shape of the
island edge. Hence, the factor $2/\sqrt{3}$ is only an estimate, which
gives an impression on the influence of the lattice correction to the
coefficient $\kappa_1$. Our comparison with the simulation results in
Fig.~\ref{taun-fig} yields $\alpha_{\rm cont}\!\cong\!2\alpha$, i.e.\ 
$\kappa_1\!\cong\!4(1/2)^2\!=\!1$. We note that in general lattice
corrections have always to be included in a continuum description
after the effective Schwoebel barrier (for the lattice) has been
calculated from the microscopic barriers (see the comment in
ref.~\onlinecite{effes-comm}).

\end{multicols}


\begin{references}
\vspace*{-1cm}

\bibitem{Venables/etal:1984} J.~A.~Venables, G.~D.~T.~Spiller, and
M.~Hanb\"ucken, Rep.~Prog.~Phys. {\bf 47}, 399 (1984).
  
\bibitem{Volmer/Weber:1926} M.~Volmer and A.~Weber,
Z.~Phys.~Chem.~(Leipzig) {\bf 119}, 277 (1926).

\bibitem{Stranski/Krastanov:1938} I.~N.~Stranski and L.~Krastanov,
Sitzungsber.~Akad.~Wiss. Wien {\bf 146}, 797 (1938).

\bibitem{Bauer/Merve:1986} E.~Bauer and J.~H.~v.~d.~Merwe,
Phys.~Rev.~B {\bf 33}, 3657 (1986).

\bibitem{Brune:1998} H.~Brune, Surf.~Sci.~Rep. {\bf 31}, 121 (1998).

\bibitem{Politi/etal:2000} P.~Politi, G.~Grenet, A.~Marty, A.~Ponchet
  and J.~Villain, Phys. Rep. 324, 271 (2000).

\bibitem{Tersoff/etal:1994} J.~Tersoff, A.~W.~Denier van der Gon, and
  R.~M.~Tromp, Phys.~Rev.~Lett. {\bf 72}, 266 (1994).
  
\bibitem{Ehrlich/Hudda:1966+Schwoebel:1969} G.~Ehrlich and
  F.~G.~Hudda, J.~Chem.~Phys. {\bf 44}, 1039 (1966); R.~L.~Schwoebel,
  J.~Appl.~Phys.  {\bf 40}, 614 (1969).
  
\bibitem{Zhang+Lagally:1994} Z.~Zhang and M.~G.~Lagally,
  Phys.~Rev.~Lett. {\bf 72}, 693 (1994); Z.~Zhang and M.~G.~Lagally,
  Science {\bf 276}, 377 (1997).

\bibitem{DasSarma/etal:1999} S.~Das Sarma, P.~Punyindu, and
  Z.~Toroczkai, e-print cond/mat~9908013.

\bibitem{Bromann/etal:1995} K.~Bromann, H.~Brune, H.~R\"oder, and
  K.~Kern, Phys.~Rev.~Lett. {\bf 75}, 677 (1995).

\bibitem{Meyer/etal:1995} J.~A.~Meyer, J.~Vrijmoeth, H.~A.~van der Vegt,
E.~Vlieg, and R.~J.~Behm, Phys.~Rev. B {\bf 51}, R14790 (1995).

\bibitem{Smilauer/Harris:1995} P.~\v Smilauer and S.~Harris, Phys.~Rev. B {\bf
    51}, R14798 (1995).
  
\bibitem{Markov:1996} I.~Markov, Phys.~Rev.~B {\bf 54}, 17930 (1996).
  
\bibitem{Roos+Tringides:1998} K.~R.~Roos and M.~C.~Tringides, Surface Review
  and Letters {\bf 5}, 833 (1998).

\bibitem{Rottler/Maass:1999} J.~Rottler and P.~Maass, Phys.~Rev.~Lett.
  {\bf 83}, 3490 (1999)
  
\bibitem{Krug/etal:2000} J.~Krug, P.~Politi, and T.~Michely,
  Phys.~Rev.~B, scheduled for issue on May 15, see also e-print
  cond/mat~9912410.
  
\bibitem{Elkinani/Villain:1994} I.~Elkinani and J.~Villain, J.~Phys. I
  France {\bf 4}, 949 (1994).

\bibitem{Havlin/Ben-Avraham:1987} S.~Havlin and D.~Ben Avraham,
  Adv.~Phys. {\bf 36}, 695 (1987).

\bibitem{Barabasi/Stanley:1995} A.-L.~Barab\'asi and
  H.~E.~Stanley, {\it Fractal Concepts in Surface Growth} (Cambridge
  University Press, 1995).
  
\bibitem{Plischke/Racz:1984} M.~Plischke and Z.~R\'acz,
  Phys.~Rev.~Lett. {\bf 53}, 415 (1984).

\bibitem{Feibelman:1998} Ab initio calculation for microscopic step
  edge barriers have been performed e.g.\ by P.~J.~Feibelman,
  Phys.~Rev.~Lett. {\bf 81}, 168 (1998).
  
\bibitem{effes-comm} For the calculation of the effective step edge
  barrier $\Delta E_{\rm S}$ that determines the total escape rate
  from an island, one should average over the local escape rates
  $\Gamma_\mu\propto\exp[-\Delta E_\mu/k_{\rm B}T]$ associated with
  the set of microscopic barriers $\{\Delta E_\mu\}$ along the island
  edge. Accordingly, $\Delta E_{\rm S}=-k_{\rm
    B}T\ln\langle\exp[-\Delta E_\mu/k_{\rm B}T]\rangle$, where
  $\langle\ldots\rangle$ denotes an average over the $\Delta E_\mu$
  (with an appropriate weighting by the probabilities of an adatom to
  be at the corresponding boundary sites). Note that for fractal
  island boundaries this averaging may result in an effective $\Delta
  E_{\rm S}$ that depends on $R$.

\bibitem{Amar/Family:1996} J.~G.~Amar and F.~Family, Phys.~Rev. B {\bf
    54}, 14742 (1996); J.~G.~Amar and F.~Family, Thin Solid Films {\bf
    272}, 208 (1996).
    
\bibitem{voronoi-comm} For the definition of the Voronoi cell, see
  e.g.\ J.~M.~Ziman, {\it Models of Disorder}, (Cambridge University
  Press, 1979); In principle one should define the ``dividing lines''
  in the Voronoi construction with respect to the island {\it
    boundaries}, which would lead to curved perimeters of capture
  areas.
  
\bibitem{nucleus-comm} The critical nucleus should be defined more
  precisely by referring to a particular atomic configuration. For
  simplicity we do not distinguish between the various possible
  configurations for a given island size.
  
\bibitem{Brune/etal:1999} H.~Brune, G.~S.~Bales, J.~Jacobsen,
  C.~Boragno, and K.~Kern, Phys.~Rev.~B {\bf 60}, 5991 (1999).

\bibitem{Tang/etal:1991} S.~L.~Tang, P.~F.~Carcia, D.~Coulman, and
A.~J.~McGhie, Appl.~Phys.~Lett. {\bf 59}, 2898 (1991).

\bibitem{Bartelt/Evans:1992} M.~C.~Bartelt and J.~W.~Evans,
  Phys.~Rev.~B {\bf 46}, 12675 (1992).
  
\bibitem{Bales/Chrzan:1994} G.~S.~Bales and D.~C.~Chrzan, Phys.~Rev.~B
  {\bf 50}, 6057 (1994).

\bibitem{Amar/Family:1994} J.~G.~Amar and F.~Family, Phys.~Rev.~Lett.
  {\bf 74}, 2066 (1994).
  
\bibitem{Roeder/etal:1993} H.~R\"oder, H.~Brune, J.~P.~Bucher, and
  K.~Kern, Surf.~Sci. {\bf 298}, 121 (1993); H.~Brune, C.~Romainczyk,
  H.~R\"oder, and K.~Kern, Nature {\bf 369}, 469 (1994).
  
\bibitem{Michely/etal:1993} T.~Michely, M.~Hohage, M.~Bott, and
  G.~Comsa, Phys.~Rev.~Lett. {\bf 70}, 3943 (1993).
  
\bibitem{continuum-comm} To simplify notation, we do not distinguish
  between the diffusion coefficient $D_{\rm cont}$ in the continuum
  limit and the coefficient $D$ defining the jump rate $D/a^2$ of the
  adatoms on the substrate lattice. In two dimensions $D_{\rm
    cont}\!=\!D/4$. In the case of heteroepitaxy, moreover, one has to
  replace $D/a^2$ by the jump rate on top of an island.  Also, we do
  not introduce the edge crossing probability $\alpha_{\rm
    cont}\!=\!\kappa_\alpha\alpha$ in the continuum limit, where the
  correction factor $\kappa_\alpha$ depends on both the lattice type
  and the shape of the island boundary. For a discussion of the
  lattice corrections, see Appendix~\ref{taun-app}.

\bibitem{Bartelt/Evans:1999} M.~C.~Bartelt and J.~W.~Evans, Surface Science
  {\bf 423}, 189 (1999); M.~C.~Bartelt and J.~W.~Evans, Surf.~Sci.~Lett.  {\bf
    314}, L829 (1994).
  
\bibitem{Bott/etal:1996} M.~Bott, M.~Hohage, M.~Morgenstern,
  T.~Michely, and G.~Comsa, Phys.~Rev.~Lett. {\bf 76}, 1304 (1996).

\bibitem{Amar/etal:1994} J.~G.~Amar, F.~Family, and P.-M.~Lam,
  Phys.~Rev. B {\bf 50}, 8781 (1994).

\bibitem{Liu/etal:1993} S.~Liu, Z.~Zhang, G.~Comsa, and H.~Metiu,
  Phys.~Rev.~Lett. {\bf 71}, 2967 (1993).
  
\bibitem{Bartelt/Evans:1995} M.~C.~Bartelt and J.~W.~Evans,
  Phys.~Rev.~Lett. {\bf 75} 4250 (1995).
  
\bibitem{alpha-comm} Writing $D/6a^2\!=\!\nu\exp(-E_0/k_{\rm B}T)$ for
  the in-layer hopping rate and $\alpha D/6a^2\!=\!\nu'\exp(-E_{\rm
    S}/k_{\rm B}T)\!=\![(\nu'/\nu)\exp(-\Delta E_{\rm S}/k_{\rm
    B}T)](D/6a^2)$ for the hopping rate over the step edge, one should
  more generally take $\alpha=(\nu'/\nu)\exp(-\Delta E_{\rm S}/k_{\rm
    B}T)$, if the attempt frequencies $\nu$ and $\nu'$ are different.
  
\bibitem{Ratsch/etal:1994} C.~Ratsch, A.~Zangwill, P.~\v Smilauer, and
  D.~D.~Vvedensky, Phys.~Rev.~Lett. {\bf 72}, 3194 (1994).
  
\bibitem{Wang/Ehrlich:1991} S.-C.~Wang and G.~Ehrlich,
  Phys.~Rev.~Lett. {\bf 67}, 2509 (1991).
  
\bibitem{Stumpf/Scheffler:1994} R.~Stumpf and M.~Scheffler,
  Phys.~Rev.~Lett. {\bf 72}, 254 (1994).
  
\bibitem{Jacobsen/etal:1995} J.~Jacobsen, K.~Jacobsen, P.~Stoltze, and
  J.~N{\o}rskov, Phys.~Rev.~Lett. {\bf 74}, 2295 (1995).

\bibitem{Stroscio/Pierce:1994} J.~A.~Stroscio and D.~T.~Pierce,
Phys.~Rev. B {\bf 49}, 8522 (1994).
  
\bibitem{Mulheran/Robbie:2000} P.~A.~Mulheran and D.~A.~Robbie,
  Europhys.~Lett. {\bf 49}, 617 (2000).
  
\bibitem{Harris:1995} S.~Harris, Phys.~Rev.~B {\bf 52}, 16793 (1995).

\bibitem{om-mf-comm} It is possible to include this spatial dependence of the
  diffusion profile into the stochastic description but the treatment then
  becomes comparable in complexity with a direct Monte Carlo simulation of the
  second layer nucleation process.
  
\bibitem{gamma-comm} If $\Gamma=D/Fa^4\lesssim1$ one would deal with
  a physical situation not considered in this work, where the mobility
  of adatoms is irrelevant.
  
\bibitem{Kallabis/etal:1998} H.~Kallabis, P.~L.~Krapivsky, and
  D.~E.~Wolf, Eur.~Phys.~J B {\bf 5}, 801 (1998).

\bibitem{Shi/etal:1996} Z.-P.~Shi, Z.~Zhang, A.~K.~Swan, and
  J.~F.~Wendelken, Phys.~Rev.~Lett. {\bf 76}, 4927 (1996).

\bibitem{prefactor-comm} For $i\!=\!2$ the various prefactors entering
  the transition rates displayed in Fig.~\ref{rate-illus1-fig} are
  $\kappa_{\rm a,1}^{(2)}=0.9$, $\kappa_{\rm a,1}^{(3)}=0.3$,
  $\kappa_{\rm a,2}^{(3)}=3$, $\kappa_{\rm a,1}^{(4)}=11$,
  $\kappa_{\rm a,2}^{(4)}=3$, and $\kappa_{\rm d,2}=0.5$. The
  relatively large factor $\kappa_{\rm a,1}^{(4)}=11$ may be
  understood as effectively taking into account the additional
  pathways resulting from states with $n\!>\!n_\star\!=\!4$. For
  $i\!=\!1$ we found $\kappa_{\rm a,1}^{(2)}=1$ to give an optimal fit
  for a reduced diagram, in which only the states with $n\!\le\!2$
  from Fig.~\ref{rate-illus1-fig} are taken into account .

\bibitem{Kyuno/Ehrlich:1998} K.~Kyuno and G.~Ehrlich, Phys.~Rev.~Lett.
  {\bf 81}, 5592 (1998).

\bibitem{ag-comm} The attempt frequency $\nu'$ for the jumps over the
  step edge (see the comment in ref.~\onlinecite{alpha-comm}), however,
  turned out to be unphysically large. When assuming the attempt
  frequency $\nu$ for in-layer jumps to be the same as $\nu'$, the
  experimental data could be fitted by allowing for a very slight
  decrease of $\Delta E_{\rm S}$ with increasing temperature from
  $\Delta E_s\!\approx\!0.11{\rm eV}$ for $T\!=\!120{\rm K}$ to
  $\Delta E_s\!\approx\!0.10{\rm eV}$ for $T\!=\!130{\rm K}$.
  
\bibitem{Kalff/etal:1998} M.~Kalff, G.~Comsa, and Th.~Michely,
  Phys.~Rev.~Lett.  {\bf 81}, 1255 (1998).

\bibitem{Vegt/etal:1992} H.~A.\ van der Vegt, H.~M.\ van Pinxteren,
M.~Lohmeier, E.~Vlieg, and J.~M.~C.\ Thornton, Phys.~Rev.~Lett. {\bf
  68}, 3335 (1992).

\bibitem{Vrijmoeth/etal:1994} J.~Vrijmoeth, H.~A.\ van der Vegt,
  J.~A.~Meyer, E.~Vlieg, and R.~J.~Behm, Phys.~Rev.~Lett. {\bf 72},
  3843 (1994).

\bibitem{Kunkel:1990} R.~Kunkel, B.~Poelsema, L.~K.~Verheij, and
  G.~Comsa, Phys.~Rev.~Lett. {\bf 65}, 7333 (1990).
  
\bibitem{Bott/etal:1992} M.~Bott, T.~Michely, and G.~Comsa, Surf.~Sci.
  {\bf 272}, 161 (1992).
  
\bibitem{Smilauer:1993} P.~\v Smilauer, M.~R.~Wilby, and
  D.~D.~Vvedensky, Phys.~Rev.~B {\bf 47}, 4119 (1993).

\bibitem{Ernst/etal:1994} H.~J.~Ernst, F.~Fabre, R.~Folkerts, and
  J.~Lapujoulade, Phys.~Rev.~Lett. {\bf 72}, 112 (1994).
  
\bibitem{Kalff/etal:1999} M.~Kalff, P.~\v Smilauer, G.~Comsa and
  T.~Michely, Surf.~Sci.~Lett., {\bf 426}, L447 (1999).
  
\bibitem{Ernst/etal:2000} R.~Gerlach, Th.~Maroutian, L.~Douillard, and
  H.-J.~Ernst, to be published.

\bibitem{Michely/etal:1999} T.~Michely, M.~Kalff, J.~Krug, P.~Politi,
  Physikalische Bl\"atter {\bf 55}, No.~11, 53 (1999).

\bibitem{Jensen:1999} P.~Jensen, Rev.~of Mod.~Phys. {\bf 71}, 1695 (1999).

\end{references}
\end{document}